\documentclass[floatfix,amsmath,amssymb,aps,prb,groupedaddress,twocolumn,superscriptaddress]{revtex4-2}

\usepackage[margin=1.7cm]{geometry}
\pdfoutput=1

\usepackage{graphicx}
\usepackage{amsthm,amssymb,amsmath,braket,mathdots,mathtools}
\usepackage{bm}

\usepackage{psfrag}
\usepackage{relsize,amsbsy}
\usepackage[export]{adjustbox}
\usepackage{dsfont}
\usepackage{xcolor,tikz}
\usepackage{lipsum}
\usepackage{ragged2e}

\usepackage{mwe}
\usepackage{tabularx}
\usepackage{multirow}
\usepackage[normalem]{ulem}
\usepackage[colorlinks,bookmarks=true,citecolor=blue,linkcolor=red,urlcolor=blue]{hyperref}

\newcommand{\be}{\begin{align}}
\newcommand{\ee}{\end{align}}

\newcommand{\bit}{\begin{enumerate}}
	\newcommand{\eit}{\end{enumerate}}

\begin{document}

\title{Density of States of the lattice Schwinger model}

\author{Irene Papaefstathiou}
\affiliation{Max-Planck-Institut f{\"u}r Quantenoptik, Hans-Kopfermann-Str. 1, D-85748 Garching, Germany}
\affiliation{Munich Center for Quantum Science and Technology (MCQST), 80799 Munich, Germany}
\author{Daniel Robaina}
\affiliation{Max-Planck-Institut f{\"u}r Quantenoptik, Hans-Kopfermann-Str. 1, D-85748 Garching, Germany}
\author{J. Ignacio Cirac}
\affiliation{Max-Planck-Institut f{\"u}r Quantenoptik, Hans-Kopfermann-Str. 1, D-85748 Garching, Germany}
\affiliation{Munich Center for Quantum Science and Technology (MCQST), 80799 Munich, Germany}
\author{Mari Carmen Ba\~nuls}
\affiliation{Max-Planck-Institut f{\"u}r Quantenoptik, Hans-Kopfermann-Str. 1, D-85748 Garching, Germany}
\affiliation{Munich Center for Quantum Science and Technology (MCQST), 80799 Munich, Germany}

\date{\today}

\begin{abstract}
Using a recently introduced tensor network method, we study the density of states of the lattice Schwinger model, a standard testbench for lattice gauge theory numerical techniques, but also the object of recent experimental quantum simulations. We identify regimes of parameters where the spectrum appears to be symmetric and displays the expected continuum properties even for finite lattice spacing and number of sites. However, we find that for moderate system sizes and lattice spacing of $ga\sim O(1)$, the spectral density can exhibit very different properties with a highly asymmetric form. We also explore how the method can be exploited to extract thermodynamic quantities.
\end{abstract}

\maketitle

\section{Introduction}

Interacting quantum many-body systems represent a challenge for analytical and numerical methods, yet they are key to the understanding of many fundamental physical phenomena. 
For this reason, and because most of the interesting systems are not exactly solvable, considerable effort is devoted to the development of very different numerical techniques to address these problems. Prominent examples include Monte Carlo algorithms~\cite{Metropolis1953} and tensor network (TN) methods~\cite{schollwoeck2011,verstraete2008,orus2014}.

One of the most significant properties of a quantum many-body problem is the density of states (DOS). Knowing the distribution of energy eigenstates gives access to the partition function and consequently to all the thermodynamic properties of the system.  In the context of lattice gauge theories (LGT), 
approximating the DOS has been proposed~\cite{langfeld2012density} as a method to overcome the sign problem~\cite{Troyer2005} that appears, for instance, in the presence of a finite chemical potential.
But computing the DOS of a many body problem in general, or a LGT in particular, is a difficult task.

Several approximate methods have been devised to address this question.
For classical statistical models broad histogram methods exist, as the Wang-Landau algorithm~\cite{Wang2001,Trebst2004}, and they have inspired variants that can be used for some quantum problems~\cite{Troyer2003,Wessel2007}.
A different technique is the kernel polynomial method (KPM)~\cite{weisse2006kernel}. Based on a Chebyshev expansion of the Dirac delta function, it is widely used in the single-particle quantum scenario. Only recently the application of tensor network algorithms to this problem has been explored~\cite{schrodi2017density,yang2020probing}.
The advantage of TN methods is that, in principle, they are free from the sign problem, and thus suitable for scenarios out of reach for Monte Carlo based algorithms. 
The method introduced in~\cite{yang2020probing} precisely adapts the KPM method to an interacting scenario where the Hamiltonian can be expressed as a matrix product operator (MPO)~\cite{verstraete2004,pirvu2010,zwolak2004} and can be used to estimate DOS, microcanonical averages and other spectral properties, but also to probe thermalization. 

 Specific methods have been proposed in the particular context of LGT. They include the Linear Logarithmic Relaxation method (LLR)~\cite{langfeld2012density,langfeld2016efficient}, based on the Wang-Landau algorithm, which was introduced to improve the precision in the calculation of the DOS, with the specific aim of determining observables at finite density lattice field theories, and the Functional Fit Approach (FFA)~\cite{gattringer2015density, giuliani2016developing, giuliani2017density} and have been used in a variety of models, such as 
compact $U(1)$, $SU(2)$ and $SU(3)$ LGT~\cite{langfeld2012density}, $SU(2)$ gauge theory at finite densities with heavy quarks~\cite{langfeld2013two} and the $\mathds{Z}_{3}$ spin model at finite density, where the sign problem is present~\cite{langfeld2014density}. These methods rely on Monte Carlo techniques and therefore present challenges when applied to systems with dynamical fermions~\cite{gattringer2019, Gattringer:2019egx}. 

 In this work, we explore the performance of the TN method introduced in~\cite{yang2020probing} in a $U(1)$ LGT. More concretely, we consider the lattice Schwinger model, which includes gauge and dynamical fermionic degrees of freedom. 
 The features of the Schwinger model, one of the simplest LGT that nonetheless exhibits non-trivial phenomena common to more complex theories, have established it as a usual testbench for lattice techniques. It has also been the object of the first experimental quantum simulation of a LGT using trapped ions~\cite{martinez2016real}, and efficient representations of the model have been suggested for its study in a quantum computer~\cite{PhysRevA.98.032331}. More recently also a closely related model with $U(1)$ gauge symmetry has been experimentally realized with ultracold atoms~\cite{Yang:2020yer}.
 Many properties of the Schwinger model, including its mass spectrum, thermal equilibrium properties and dynamics, have been systematically investigated using TN approximations for the relevant states~\cite{banuls2020review,qtflag2020}. These studies have demonstrated the suitability of TN to describe the physically relevant states in LGT scenarios, and to attain precise continuum extrapolations. 
 However, the standard techniques used in such studies do not provide easy access to high energy eigenstates, or to the spectral properties at energy densities far from the edges of the spectrum.
 
 Using the above mentioned technique, allows us to directly study the DOS of the lattice model. 
 In particular, we examine the dependence of the DOS with system size and lattice spacing, parameters that need to be varied in the (classical or quantum) simulation in order to approach the continuum limit. We find that the shape of the DOS changes dramatically between a \emph{lattice-dominated} regime, where it is highly asymmetrical and exhibits sharp features, and a \emph{continuum-like} regime, in which it resembles a Gaussian distribution. The last observation is in accordance with the results of~\cite{elser1996qed}, to our knowledge the only previous study of the DOS of the model.
 As a second goal, we explore the potential of the DOS approximation to calculate different observables in the canonical ensemble, in the spirit of the LLR method, and what are the limitations of this method in comparison to directly approximating thermal states with TN.

 The rest of the paper is organised as follows: In Sect.~\ref{sec:Model} we briefly introduce Schwinger model on the lattice with the Kogut-Susskind formulation of staggered fermions, that is then mapped to a system of long range spin-spin interactions. We then proceed with presenting the methods used in Sect.~\ref{sec:Methods} and the results follow in Sect.~\ref{sec:results}. We close with a discussion in Sect.~\ref{sec:discussion}.
 
 \section{Model}\label{sec:Model}
 
 In this paper we focus on QED in two space-time dimensions, known as the Schwinger model, and, more concretely, on its discrete lattice version.
  The Schwinger model is one of the simplest gauge theories, yet shares some of the most interesting features of quantum chromodynamics (QCD)~\cite{schwinger1962gauge, coleman1976more}, such as confinement and a broken chiral symmetry. This justifies its role as a standard testbench for LGT techniques.
  The model has also constituted a natural first target to benchmark the performance of TNS techniques for LGT. This enterprise started with a successful application of the density matrix renormalization group (DMRG) algorithm  in~\cite{Byrnes:2002nv}, and has led in more recent years to the systematic exploration of spectral properties, thermal equilibrium or dynamics of the lattice model and its continuum extrapolation (see e.g.~\cite{banuls2020review} for a review). 
Also the quantum simulation of the model has been proposed for various experimental platforms~\cite{qtflag2020}.

 Specifically, the Kogut-Susskind staggered fermion formulation~\cite{kogut1975j} of the lattice Schwinger model reads:
 \begin{align}
    H &=\frac{g^{2}a}{2}\sum_{n}L_{n}^{2}+ m\sum_{n}(-1)^{n}\Phi^{\dagger}_{n}\Phi_{n},\nonumber\\
    &-\frac{i}{2a} \sum_{n}\left(\Phi_{n}^{\dagger}e^{i\theta_{n}}\Phi_{n+1}-h.c.\right)\,,
\label{eq: Schwinger lattice}
\end{align}
with $a$ the lattice spacing, $m$ the fermion mass and $g$ the coupling constant. Operators ${\Phi}_{n}({\Phi}_{n}^{\dagger})$ annihilate (create) the (single component) fermion mode on each lattice site $n$, and satisfy canonical anticommutation relations $\{\Phi^{\dagger}_{n}, \Phi_{m} \}=\delta_{nm}$ and
$\{\Phi_{n}, \Phi_{m} \}=0$.
Gauge degrees of freedom residing on the link between sites $n$ and $n+1$ are represented by canonically conjugate operators $\theta_n$ and $L_n$, which satisfy $[\theta_{n},L_{m}]=i\delta_{nm}$ and, in the continuum limit, respectively correspond to the vector potential and the electric field.
Additionally, physical states need to satisfy the discrete version of Gauss law $L_{n}-L_{n-1}=\Phi^{\dagger}_{n}\Phi_{n}-\frac{1}{2}\left[1-(-1)^{n}  \right]$~\cite{hamer1997series}.

The Hamiltonian of Eq.~\eqref{eq: Schwinger lattice} can be mapped to a spin model via the Jordan-Wigner transformation 
$\Phi_{n}=\prod_{k<n}(i\sigma^{z}_{k})\sigma^{-}_{n}$ with $\sigma^{\pm}=\frac{1}{2}(\sigma^{x}\pm i\sigma^{y})$, where $\sigma^{\alpha}$, for $\alpha=x,\ y,\ z$, are the Pauli matrices. 
Additionally, for a system with open boundary conditions, it is possible to explicitly solve Gauss law,
which results in a spin chain with long-range interactions~\cite{banks1976strong,hamer1997series}.
Usually, the Hamiltonian is multiplied by a factor $\frac{2}{g^2 a}$ to result in an adimensional operator which, 
for a chain of $N$ sites, reads
 \begin{align}
    W&:= \frac{2}{g^2 a} H = x\sum_{n=0}^{N-2}\left[\sigma_{n}^{+}\sigma_{n+1}^{-}+\sigma_{n}^{-}\sigma_{n+1}^{+}\right]\nonumber\\
    &+\frac{\mu}{2}\sum_{n=0}^{N-1}\left[1+(-1)^{n}\sigma^{z}_{n}\right]\nonumber\\ 
    &+\sum_{n=0}^{N-2}\left[ l+\frac{1}{2}\sum_{k=0}^{n}((-1)^{k}+\sigma_{k}^{z}) \right]^{2},
    \label{eq: Schwinger spin}
\end{align}
where the relevant (adimensional) parameters are now $x=\frac{1}{g^{2}a^{2}}$ 
and $\mu=\frac{2m}{g^{2}a}$, while $l$ represents the background field. 
Written in this form the model is suitable to be studied numerically with TNS methods.

\section{Methods}\label{sec:Methods}

In order to compute the density of states of the model \eqref{eq: Schwinger spin} we use the TNS techniques introduced in~\cite{yang2020probing}, which we summarize here for completeness.

Given a Hamiltonian $H$ and an operator $O$ we can define a generalized density of states function
\begin{align}
    g(E;O) &= \sum_{k}\delta(E-E_{k})\langle k |O|k\rangle,
\label{gO}
\end{align}
where the sum runs over all energy eigenstates $\ket{k}$ of the Hamiltonian, $H\ket{k}=E_k\ket{k}$.
Notice that, up to a normalization factor, the usual density of states corresponds to the function for the identity operator  $g(E;\mathds{1})$.

The functions~\eqref{gO} can be approximated by a finite sum of $M$ Chebyshev polynomials (see appendix~\ref{Appendix A}), 
 based in the corresponding approximation of the Dirac delta function $\delta(x)\approx \delta_M(x)$. Explicitly, 
\begin{align}\label{Eq:delta_expansion}
    \delta_{M} (x-x_0)= \frac{1}{\pi \sqrt{1-x^{2}}} \sum_{n=0}^{M-1} 
    (2-\delta_{n 0})\gamma_{n}^{M}
    T_{n}(x_0)T_{n}(x),
\end{align}
with the coefficients $\gamma_{n}^{M}$ (explicitly shown in appendix~\ref{Appendix A}) corresponding to the Jackson kernel in the kernel polynomial method~\cite{weisse2006kernel}.

Specifically we define
 \begin{align}
g_{M}(E;O) \equiv \sum_{k} \delta_{M}(E-E_{k})\langle k |O|k\rangle, \label{eq: DOS2}
\end{align}
and by substituting Eq.~\eqref{Eq:delta_expansion} we get
 \begin{align}
    g_M(E;O)=   \frac{1}{\pi \sqrt{1-\Tilde{E}^{2}}} \sum_{n=0}^{M-1}(2-\delta_{n 0}) {\gamma}^{M}_{n}\mu_{n}(O)T_{n}(\Tilde{E}),  \label{eq:gM}
\end{align}
where we have defined the rescaled and shifted energy $\Tilde{E} = \alpha E+\eta$, and correspondingly the   Hamiltonian $\Tilde{H} = \alpha H+\eta$, such that the spectrum lies in the interval $[-1,1]$, and we identify the moments~\cite{yang2020probing}
\begin{align}
    \mu_{n}(O) :=\alpha \,\textrm{tr}\left(T_{n}(\Tilde{H})O\right).
\label{moments}
\end{align}

As described in \cite{yang2020probing} (see also~\cite{PhysRevB.83.195115, Wolf2015,Halimeh2015}), if the Hamiltonian is written as a matrix product operator (MPO)~\cite{PhysRevLett.93.207204,zwolak2004,pirvu2010}, one can also construct MPO approximations to its Chebyshev polynomials, starting from the exact
$T_{0}=\mathds{1}$ and $T_{1}=\Tilde{H}$, and sequentially applying the recurrence relation
$T_{n+2}(\Tilde{H})=2\Tilde{H}T_{n+1}(\Tilde{H})-T_{n}(\Tilde{H})$. The bond dimension of the resulting polynomial increases with $n$, and 
truncating it to a fixed value $D$ produces a MPO approximation  $T_{n}^{(D)}(\Tilde{H})$. Using the latter, we can estimate the moments in Eq.~\eqref{moments} for operators of interest, and thus approximate the desired functions (see~\cite{yang2020probing} for details).

The above method allows access to thermodynamic observables.
The partition function in the canonical ensemble at inverse temperature $\beta$ can be obtained from the density of states,
\begin{align}\label{partition function}
    Z(\beta) &= \int dE e^{-\beta E}g(E;\mathds{1}),
\end{align}
and, consequently, different thermodynamic quantities can be computed, such as the energy
\begin{align}
    E(\beta) &= -\frac{\partial \left(\ln Z(\beta)\right)}{\partial \beta},
\label{energy}
\end{align}
the specific heat
\begin{align} \label{eq: specific heat}
    c(\beta)&=\frac{1}{N}\frac{\partial E (\beta)}{\partial T},
\end{align}
or the entropy 
\begin{align}
   S(\beta) &= \frac{1}{T}\left[E(\beta)-F(\beta)\right],
\label{entropy}
\end{align}
with $F(\beta)=-\frac{1}{\beta}\ln Z(\beta)$ the free energy.
Also for a general observable $O$, the expectation value in the canonical ensemble can be expressed as
\begin{align}
  O(\beta) = \frac{\int dE e^{-\beta E}g(E;O)}{Z(\beta)}.
\label{canonical ensemble}
\end{align}
Using the expansion and MPO approximations $g_M(E;O)$ described above, and performing the one-dimensional integration with the Boltzmann factor $e^{-\beta E}$, we can obtain approximations for the partition function $Z_{M}(\beta)$ and the observables $O_{M}(\beta)$. For the partition function, in particular,

 \begin{align}
    Z_{M}(\beta)\propto \\ \nonumber
    & \int_{\Tilde{E}_{\min}}^{\Tilde{E}_{\max}} d \Tilde{E}  \sum_{n=0}^{M-1}(2-\delta_{n 0}) {\gamma}^{M}_{n}\mu_{n}(\mathds{1})\frac{e^{-\frac{\beta \Tilde{E}}{\alpha}}T_{n}(\Tilde{E})}{\pi \sqrt{1-\Tilde{E}^{2}}}. \label{eq:Z_M}
\end{align}
with $\tilde{E}_{\min}$ and $\tilde{E}_{\max}$ the rescaled estimates of the edges of the spectrum. The proportionality sign accounts for a factor $e^{\frac{\beta \eta}{\alpha}}/\alpha$, which is not explicitly written in the expression above.

We can approximate the functions as an alternative series using the Chebyshev expansion of the exponential~\cite{MOORE2011537}
\begin{align}
e^{bx} &= I_{0}(b)+2\sum_{n=1}^{\infty}I_{n}(b)T_{n}(x),
\end{align}
where $I_{n}(b)$ are the modified Bessel functions of order $n$.
Substituting the Boltzmann factor by this expansion in Eq.~\eqref{partition function} or the numerator of ~\eqref{canonical ensemble},   
 and using the orthogonality relation of the Chebyshev polynomials (shown in Appendix~\ref{Appendix A}),
 we can analytically integrate each term of the sum and express the result as a series of the modified Bessel functions  $I_{n}(-\beta / \alpha)$.

Specifically,
\begin{align}\label{eq: partition fuction with bessel}
    Z_M(\beta)\propto \sum_{n=0}^{M-1} N_{n}\left[(2-\delta_{n 0})\right]^{2}{\gamma}^{M}_{n}\mu_{n}(\mathds{1})I_{n}\left(-\beta/\alpha\right),
\end{align}
where $N_{n}=(1+\delta_{n 0})\pi/2$ are the norms of the polynomials.

In the same way, we get for the approximation of the numerator of Eq.~\eqref{canonical ensemble}
\begin{align}
\int dE e^{-\beta E}&g_{M}(E;O)\propto \\
&\sum_{n=0}^{M-1} N_{n}\left[(2-\delta_{n 0})\right]^{2}{\gamma}^{M}_{n}\mu_{n}(O)I_{n}\left(-\beta/\alpha\right).\nonumber
\end{align}

\section{Results}\label{sec:results}

The Hamiltonian~\eqref{eq: Schwinger spin} can be represented exactly as a matrix product operator with small bond dimension ($D=5$)~\cite{Banuls:2013ja}. This allows us to employ the method described in Sec.~\ref{sec:Methods} to approximate the density of states of the model. To the best of our knowledge, the only previous estimate of this quantity 
was performed in~\cite{elser1996qed}, using discrete spectra obtained by the method of discretized light-cone quantization (DLCQ)~\cite{pauli1993phys,  brodsky1993slac, brodsky1998quantum}. The calculation was limited to finite sizes, but the authors observed that the form of the DOS, after an extrapolation to the continuum, appeared to roughly represent a Gaussian.

It is interesting to notice that, while for local Hamiltonians it is not surprising 
that the DOS resembles a Gaussian, because it weakly converges to this shape as the system size increases~\cite{hartmann2005,keating2015}, this is not necessarily the case for the model in Eq.~\eqref{eq: Schwinger spin}, as it contains long-range interactions.
Here we explore in more detail how the DOS of the lattice model changes with the system size and the lattice spacing, and try to understand how the approximately Gaussian behavior arises as the parameters approach a regime close to the continuum. We also evaluate thermodynamic observables like the energy, the entropy, the chiral condensate and the specific heat.

\subsection{The DOS of the lattice Schwinger Model}
\label{subsec:dos}

In this work we consider exclusively the case of massless fermions $\mu=0$ and zero background field $l=0$.
We calculate the DOS of systems with system sizes between 20 and 60 sites and lattice spacing corresponding to $x$ is $1\leq x \leq 300$ (see table~\ref{systemparameters} for the precise parameters used for each size). In the following, the DOS is normalised to one, $\int dE \, \Tilde{g}(E;\mathds{1}) =1$, which we denote as $\Tilde{g}(E;\mathds{1})$.
 \begin{table}[h]
\centering
 \begin{tabular}{|c | c| c| c |c| c|} 
 \hline
N & $\text{x}_{\min}$ & $\text{x}_{\max}$\\ [0.35ex] 
\hline
20 & 1 & 40\\[0.35ex]
 \hline 
30 & 1 & 80\\[0.35ex]
 \hline 
40 & 1 & 120\\[0.35ex]
 \hline 
50 & 5 & 200\\[0.35ex]
 \hline 
60 & 5 & 300\\[0.35ex]
 \hline 
\end{tabular}
\caption{Range of lattice spacing parameters used in the calculation of the DOS for different system sizes.}
\label{systemparameters}
\end{table}

\begin{figure*}
\centering
    \includegraphics[width=0.4\linewidth]{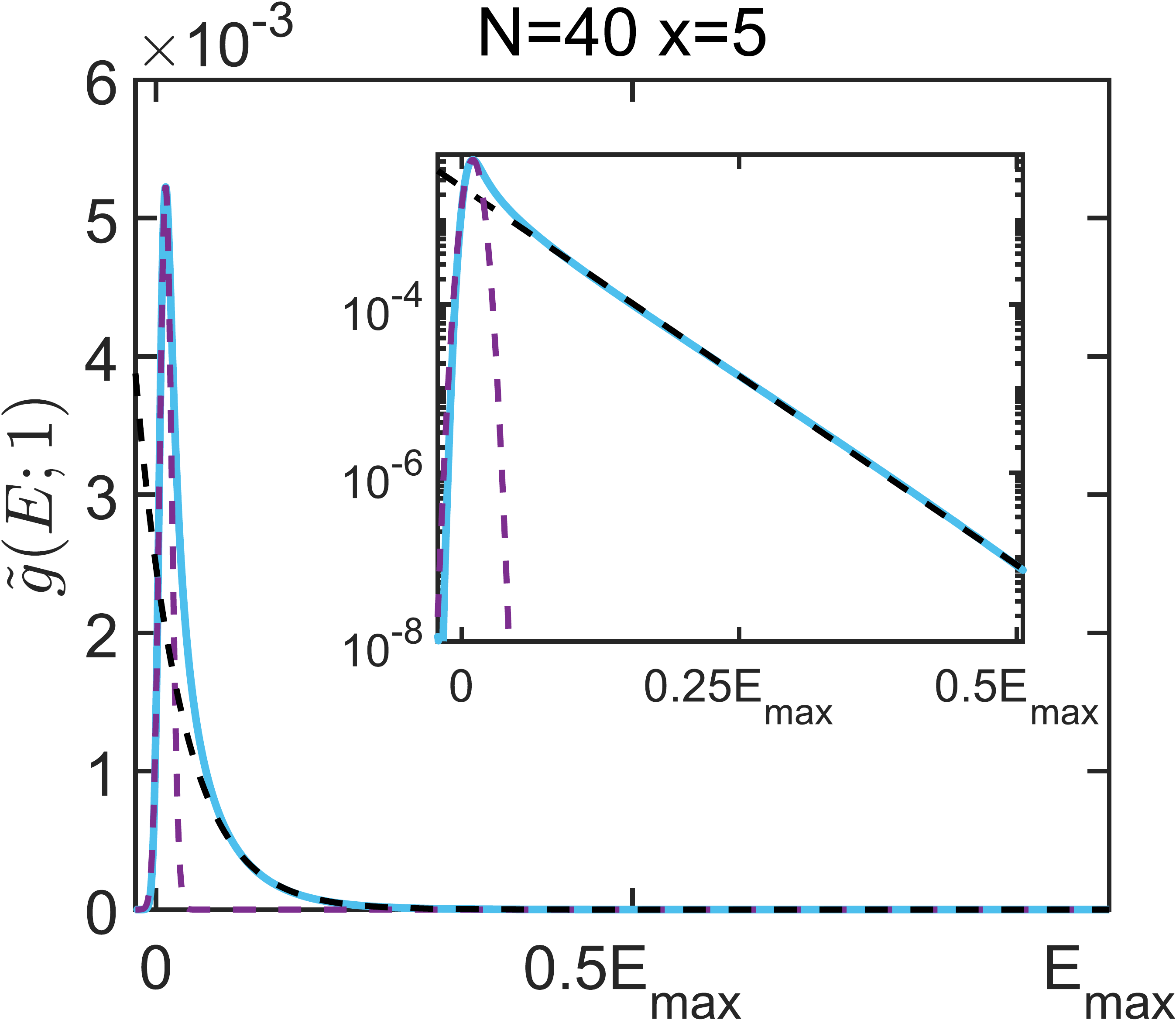}\hfil
    \includegraphics[width=0.4\linewidth]{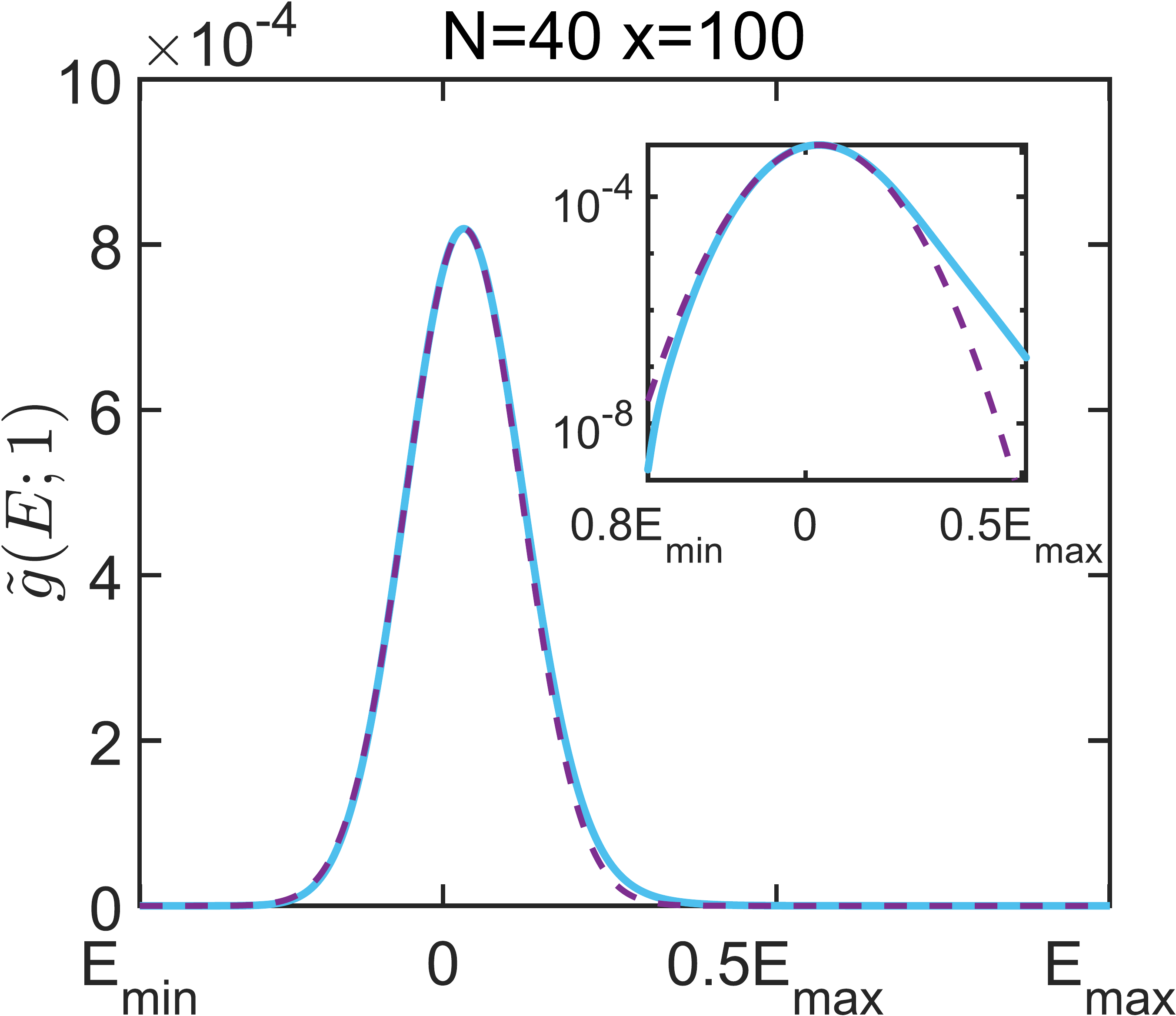}\par\medskip
    \includegraphics[width=0.4\linewidth]{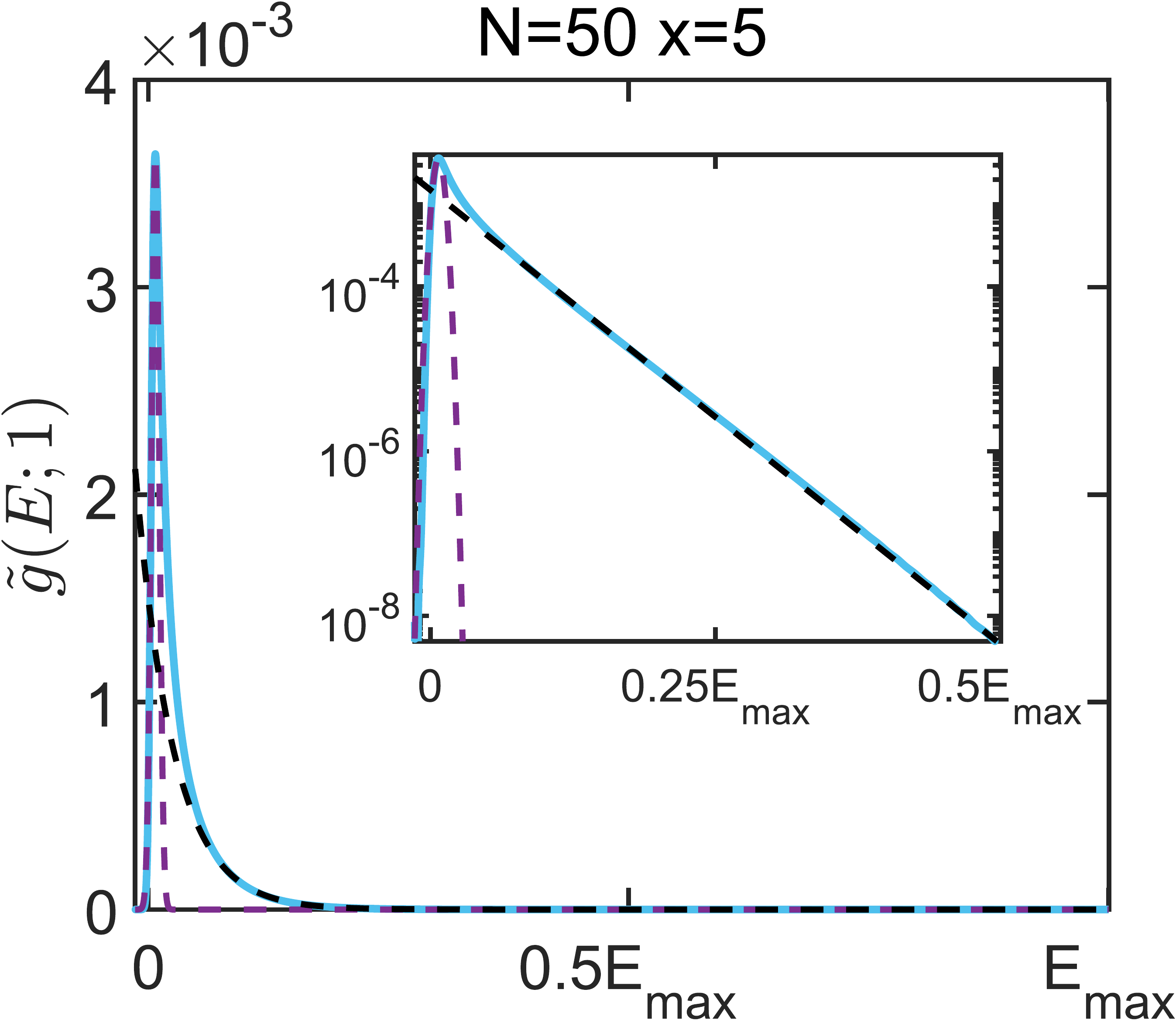}\hfil
    \includegraphics[width=0.4\linewidth]{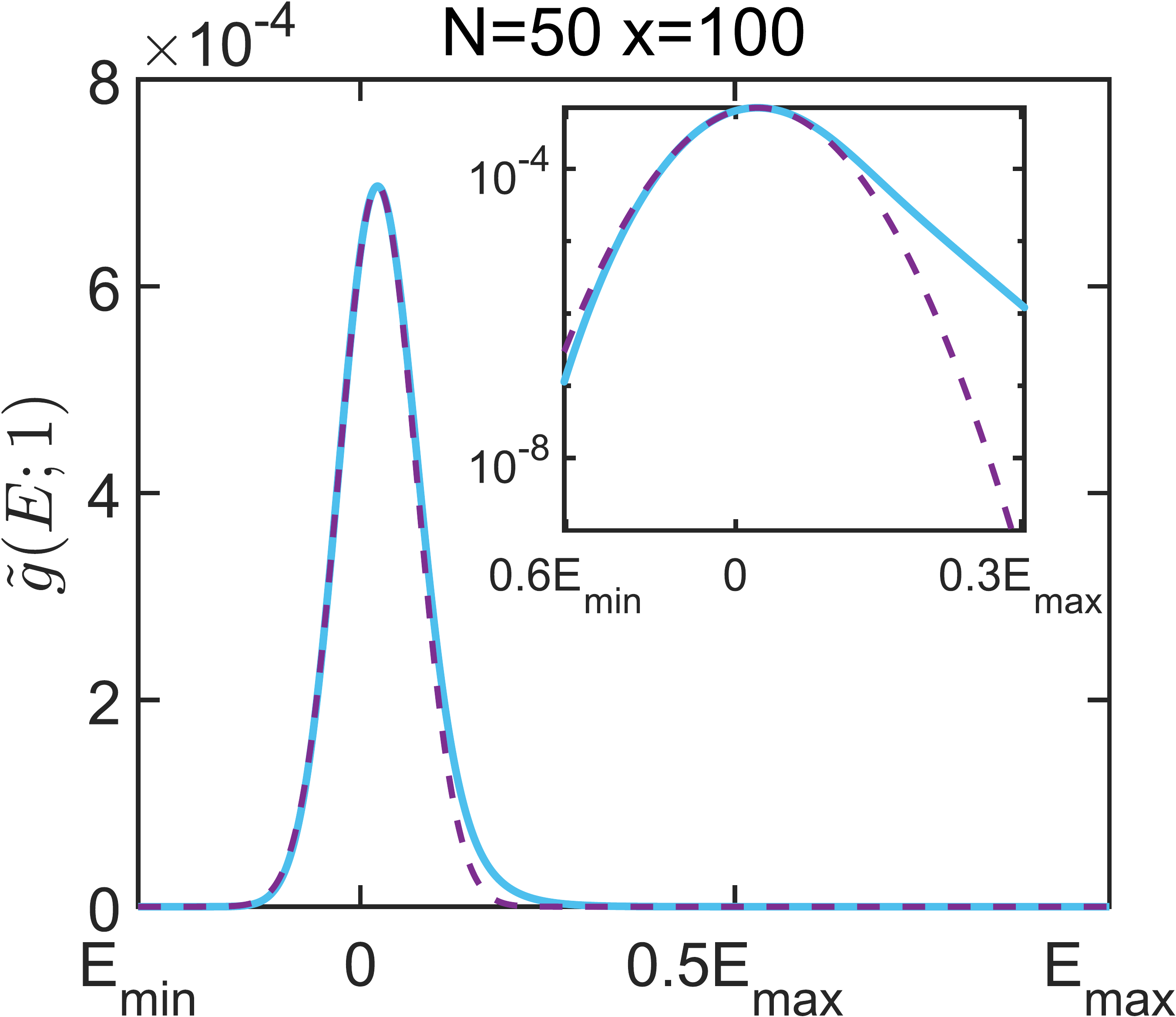}
\caption{.
Dependence of the DOS with system size and lattice spacing. The main panels show the DOS as a function of the energy in a system with $N=40$ (upper row) and $N=50$ (lower row) lattice sites for two different values of the lattice spacing parameter, $x=5$ (left column) and $x=100$ (right column). We also plot the Gaussian fit that describes the left of the peak (purple dashed lines) and (for small $x$) the exponential tails on the right (black dashed lines). The insets show a close up of the same data (with Gaussian and exponential fits) in logarithmic scale. The truncation parameters used for the simulations were $M=2000$, $D=400$ for the cases with $x=5$  and $M=500$, $D=400$ for $x=100$.
}
\label{fitting}
\end{figure*}

\begin{figure}
\centering
\includegraphics[width=0.95\columnwidth]{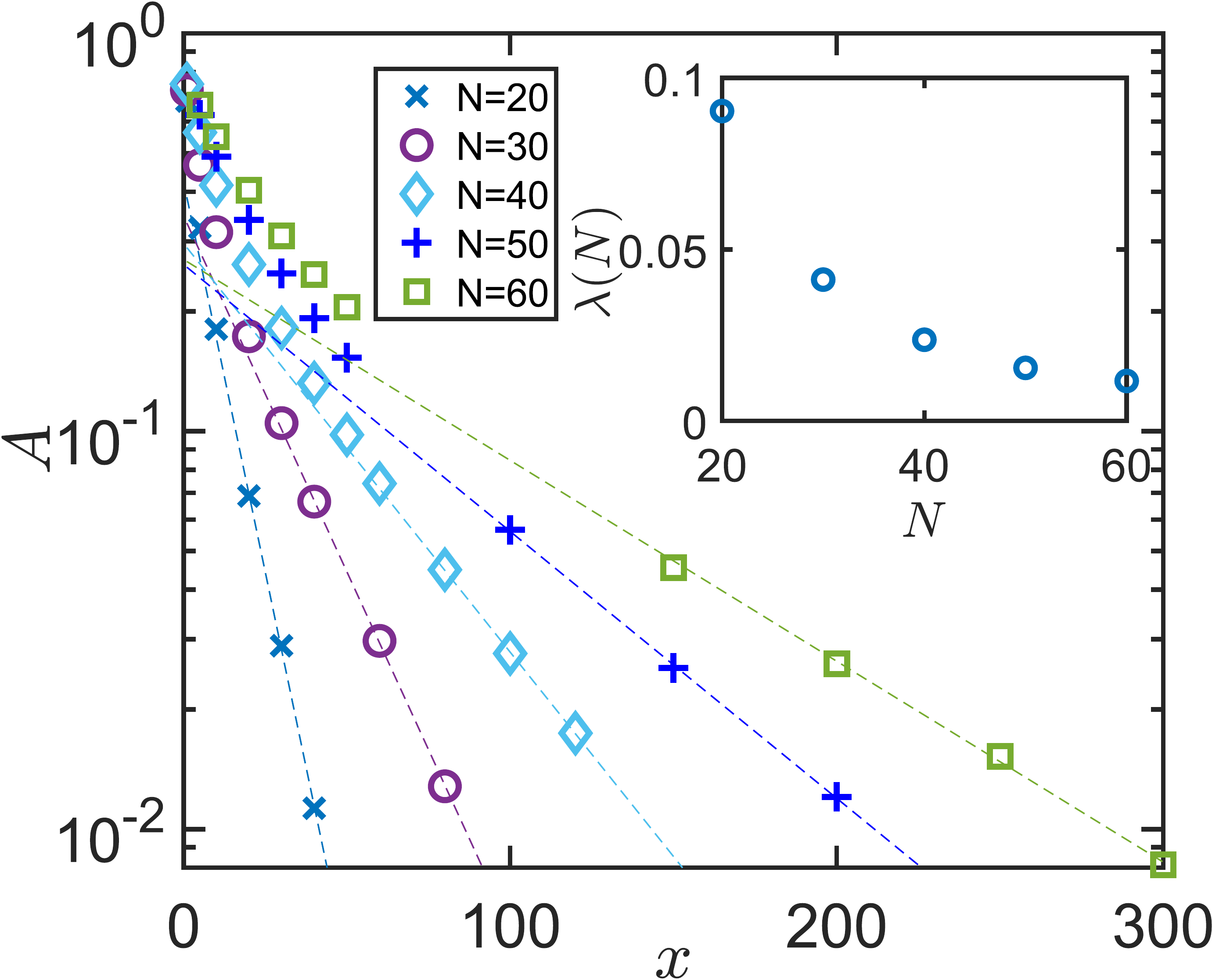}
\includegraphics[width=0.95\columnwidth]{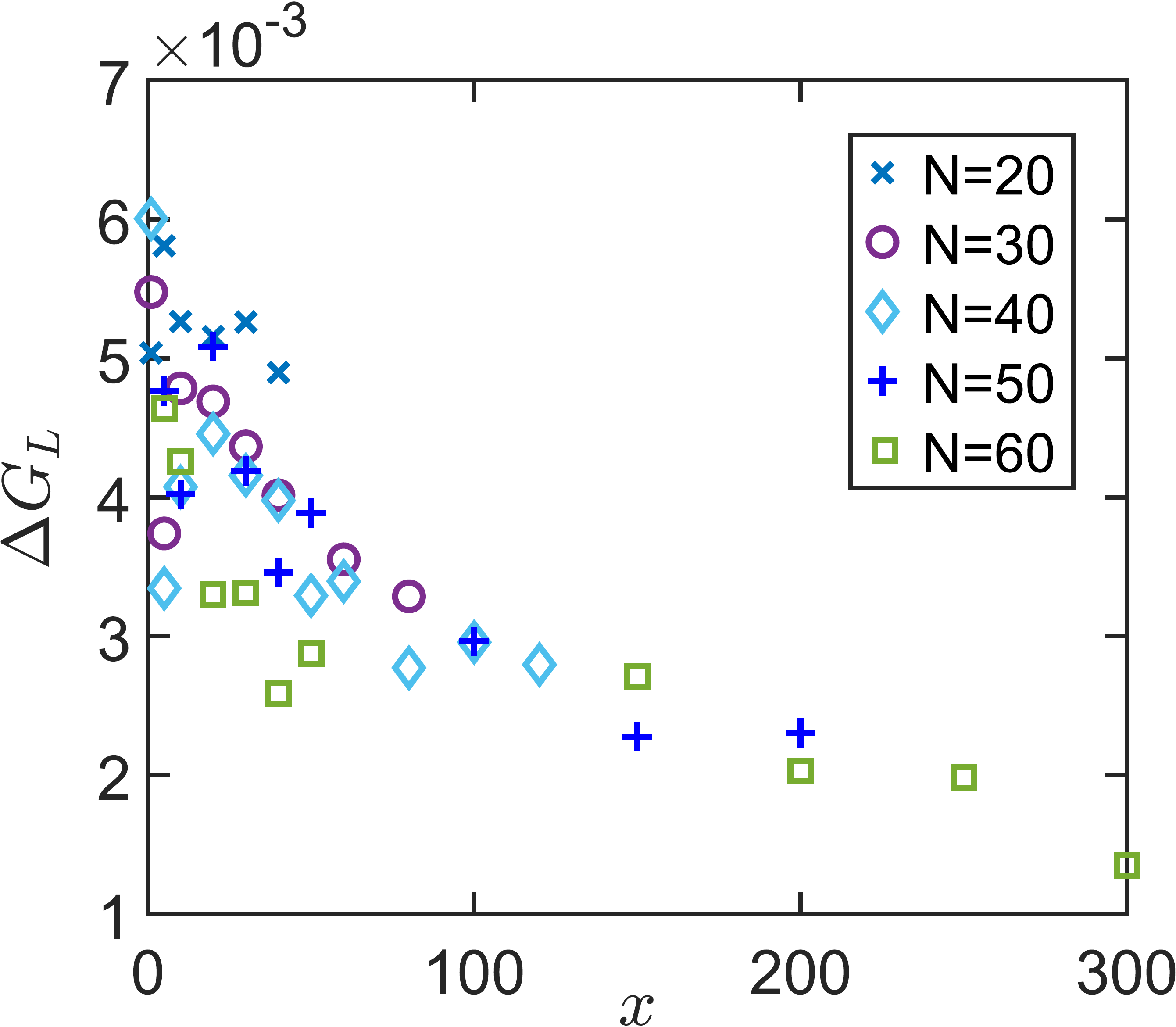}
\caption{
Asymmetry in the probability distribution, as defined in Eq.~\eqref{eq:area_diff}, as a function of the parameter $x$, for various system sizes (upper panel).
As we approach bigger values of $x$, we notice that the form of the DOS becomes symmetric. For each system size $N$, the decay of the asymmetry can be fitted to an exponential $A\propto e^{-\lambda x}$, with a faster decay for smaller systems. The resulting parameter $\lambda$ is shown in the inset as a function of the system size. The lower panel shows the  deviation with respect to a Gaussian in the area of the DOS to the left of the peak, as defined in Eq.~\eqref{areadifferenceGaussianDOS}. 
}
\label{Area difference fitting}
\end{figure}

We observe that the lattice effects dramatically affect the shape of the DOS, as shown in Fig.~\ref{fitting}. For small lattice spacing (corresponding to large values of $x$), the distribution looks similar to a Gaussian, a behaviour which was 
suggested in~\cite{elser1996qed} for the continuum limit\footnote{Notice that in~\cite{elser1996qed} the Gaussian behaviour of the DOS was only a qualitative observation, since the parameters of the numerical study did not allow approaching the continuum.}. For a fixed system size $N$, as we decrease the value of $x$ the peak becomes sharper in accordance with the behaviour suggested in~\cite{elser1996qed}. For a fixed system size $N$, as we decrease the value of $x$ the peak becomes sharper and closer to the lower edge of the spectrum, and the DOS takes a very asymmetric form, with a fast increment of the density close to the lowest energy and a much slower decrease after the peak.
We further appreciate that in the small $x$ regime (see left panels of Fig.~\ref{fitting} for $x=5$), this heavier right tail of the distribution is approximately exponential. 

In order to investigate quantitatively the deformation of the DOS, we first notice that to the left of the peak the shape of the probability distribution is still close to Gaussian. If the position of the peak is $E_p$, we can then find the Gaussian distribution that best describes the DOS for $E\in [E_{min}, E_{p}]$, namely $G(E)=Ne^{-\frac{(E-E_p)^{2}}{2\sigma^{2}}}$, i.e. the Gaussian distribution with the peak at the same position and  with variance $\sigma^2$ fixed by the half-width at half maximum. Specifically, $\sigma=(E_p-E_{1/2})/\sqrt{2 \ln 2}$, where $E_{1/2}$ is the value of the energy to the left of the peak satisfying 
$\Tilde{g}(E_{1/2};\mathds{1})=\Tilde{g}(E_{p};\mathds{1})/2$.
Finally, the normalisation constant $N$ ensures that the maximum value matches the peak of the DOS. The resulting Gaussian function is shown as a purple dashed line for each of the cases illustrated in Fig.~\ref{fitting}.
For small $x$, the right tail of the DOS decays much slower than the right tail of this Gaussian fit. Instead, an exponential decay provides a much better fit (shown in Fig.~\ref{fitting} with black dashed lines).

We can quantify the asymmetry of the distribution by computing the difference between the areas under the DOS curve to the right and to the left of the peak,
\begin{align}
A&=\int_{E_{p}}^{E_{\max}} \Tilde{g}(E;\mathds{1})\,dE-\int_{E_{\min}}^{E_{p}} \Tilde{g}(E;\mathds{1})\,dE\,.
\label{eq:area_diff}
\end{align}

This difference approaches zero as the DOS becomes closer to a symmetric distribution. The actual values of this asymmetry for a range of system sizes $N$ are shown in Fig.~\ref{Area difference fitting} as a function of  $x$. We observe that for all system sizes, the asymmetry vanishes as $x$ grows (i.e. towards small values of the lattice spacing), more slowly for larger system sizes. 
To further analyze this, we fit the last few data points for each system size, in Fig.~\ref{Area difference fitting}, to an exponential decay $A=A_0 e^{-\lambda x}$, where the parameters $A_0$ and $\lambda$ depend on $N$. The 
decay constant $\lambda$ decreases as the system size grows (see inset of Fig.~\ref{Area difference fitting}), 
consistent with a faster approach to a symmetric shape for smaller systems.

As mentioned above, to the left of the peak, the distribution is very close to a Gaussian form. To quantify this observation, we compute the difference between the area of the Gaussian $G(E)$ and the area of the DOS, to the left of the peak,
\begin{align}
  \Delta G_{L}&=\int_{E_{\min}}^{E_{p}} G(E)\,dE-\int_{E_{\min}}^{E_{p}} \Tilde{g}(E;\mathds{1})\,dE\,.  
 \label{areadifferenceGaussianDOS}
\end{align}
As we can see from the lower panel of Fig.~\ref{Area difference fitting}, for all system sizes and values of $x$ this value is very small, indicating that, as $x$ increases, the DOS approaches a symmetric distribution that resembles a Gaussian.

We note here that in the case of approaching the continuum limit, with $x\rightarrow \infty$, the Hamiltonian of Eq.~\eqref{eq: Schwinger spin} reduces to the exactly solvable XY model multiplied by a factor. In that case, there exists a unitary transformation $U$ such that
\begin{align}
    UWU^{\dagger}=-W
\end{align}
which implies a symmetric spectrum. This is not necessarily the case when $x$ is finite and the long-range interactions are present.

\begin{figure}
       \centering
\begin{minipage}[b]{0.95\columnwidth}       
\includegraphics[width=0.9\columnwidth]{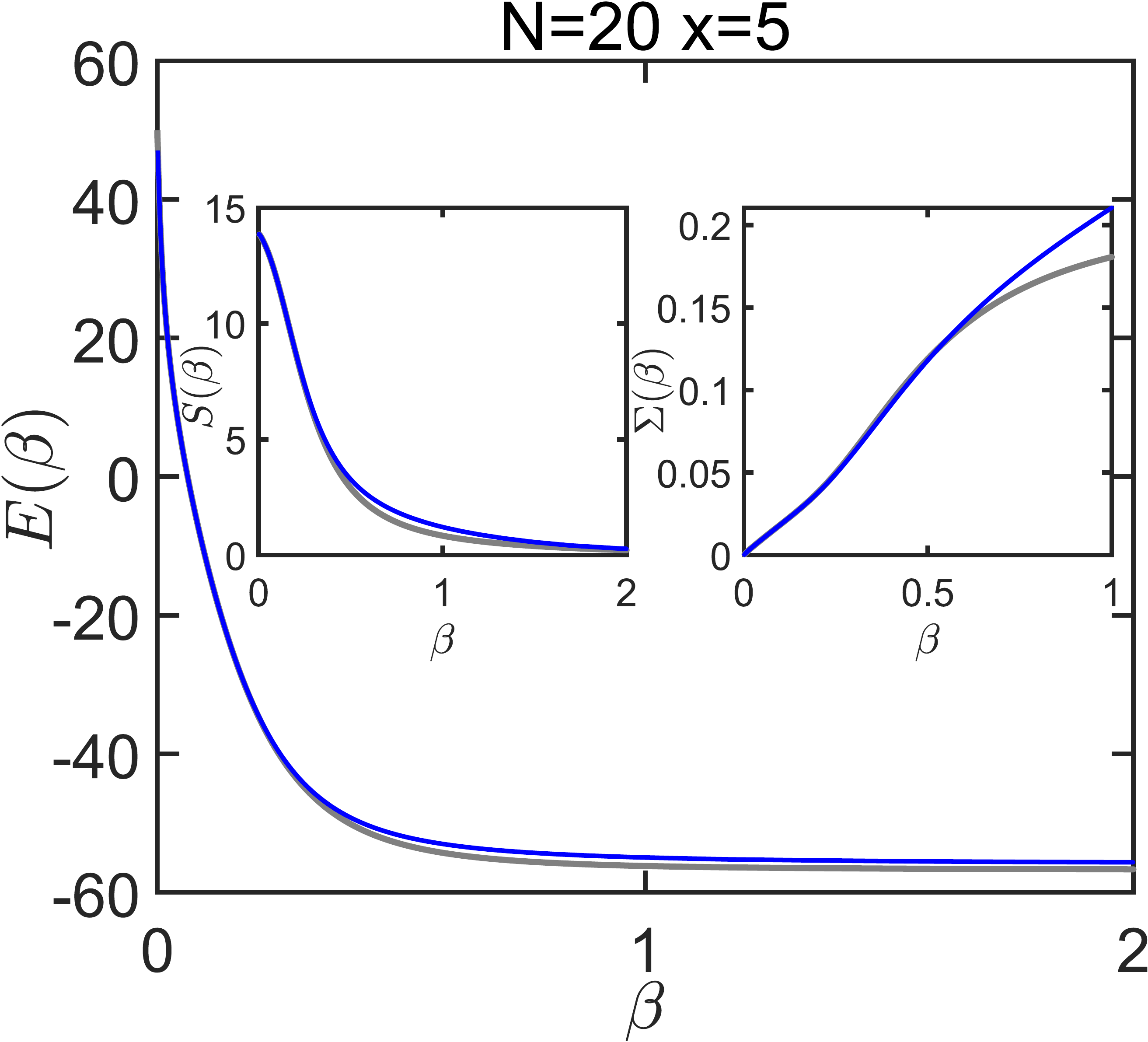}
\end{minipage}
\begin{minipage}[b]{0.95\columnwidth}       
\includegraphics[width=0.9\textwidth]{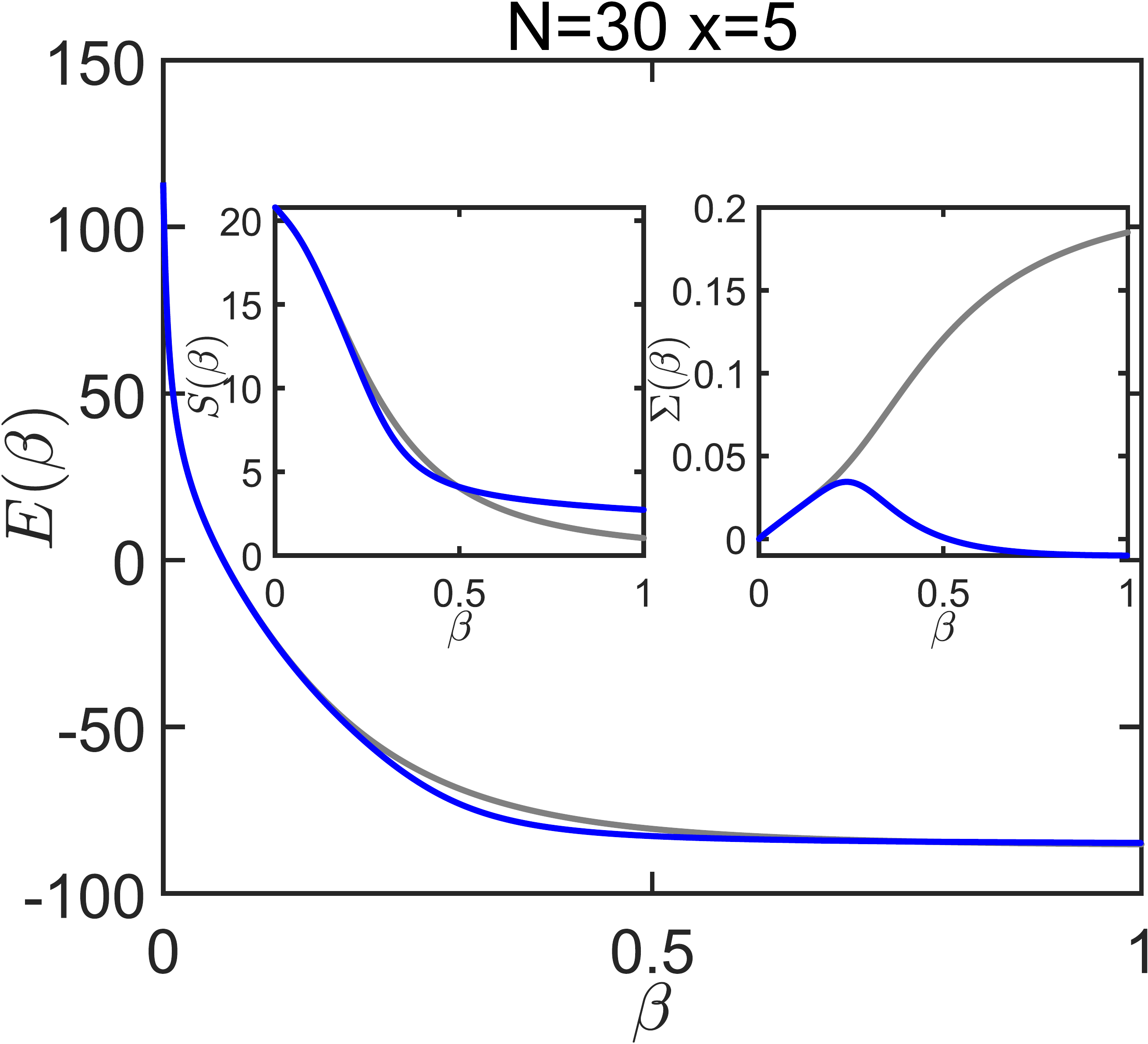}
\end{minipage}
\caption{The energy (main plot), entropy (left inset) and condensate (right inset) as a function of the inverse temperature $\beta$ in the canonical ensemble. The quantities are obtained for a system with $N=20$ sites and $x=5$, bond dimension $D=500$ and $M=2000$ (upper panel). The same quantities are calculated for a system with $N=30$ sites, $x=5$, $D=500$ and $M=2000$ (lower panel). For comparison, we show the results obtained with standard MPS techniques as described in ref.~\cite{banuls2015thermal} (grey lines).
}
\label{fig: condensateN20N30}
\end{figure}

\subsection{Physical observables}

As argued above, the DOS determines the values of thermodynamic properties, because it gives us access to the partition function and all thermal observables in the canonical ensemble through Eq.~\eqref{canonical ensemble}.
 We can thus investigate 
 how the features of the distribution observed in the previous subsection affect different thermal properties. Specifically, we compute the energy $E(\beta)$ and the entropy $S(\beta)$, as well as the value in thermal equilibrium (denoted $\Sigma(\beta)$) of the chiral condensate operator
 \begin{align}
     \hat{\Sigma}=\frac{g\sqrt{x}}{N}\sum_{n}(-1)^{n}\frac{1+\sigma_{n}^{z}}{2},
     \label{eq:condensate}
 \end{align}
     which corresponds, in the continuum limit, to the order parameter of chiral symmetry breaking. In the massless case, the temperature dependence of this parameter in the continuum limit has been solved analytically~\cite{Sachs:1991en}, while for massive fermions,
     it has been studied numerically using MPS techniques ~\cite{banuls2015thermal,banuls2016thermal,Buyens:2016ecr}.

As described in section~\ref{sec:Methods}, the partition function and thermal observables can be directly approximated as finite sums of modified Bessel functions $I_{n}(-\beta/\alpha)$, where $\alpha$ is the Hamiltonian rescaling factor, proportional to its operator norm. However, because the norm of the Hamiltonian grows with the system size (scaling as fast as $N^3$ for fixed small $x$), the argument of the functions involved, and thus the magnitude of the functions, grows fast for fixed $\beta$, leading to numerical instabilities, except for very small values of $\beta$. 
\footnote{Specifically, for $\beta_{0}=0.1$ and the system with $N=20$ and $x=5$, the Bessel function of order 10 has magnitude $|I_{10}(-\beta_{0}/\alpha)|\sim 7\times10^{15}$, whereas for the system $N=40$ and $x=5$, $|I_{10}(-\beta_{0}/\alpha)|\sim 8\times10^{129}$. }
Thus, we find it more convenient to evaluate numerically the integrals of Eq.~\eqref{partition function} and of the numerator and denominator of Eq.~\eqref{canonical ensemble}.

 \begin{figure}
\centering
    \includegraphics[width=0.9\columnwidth]{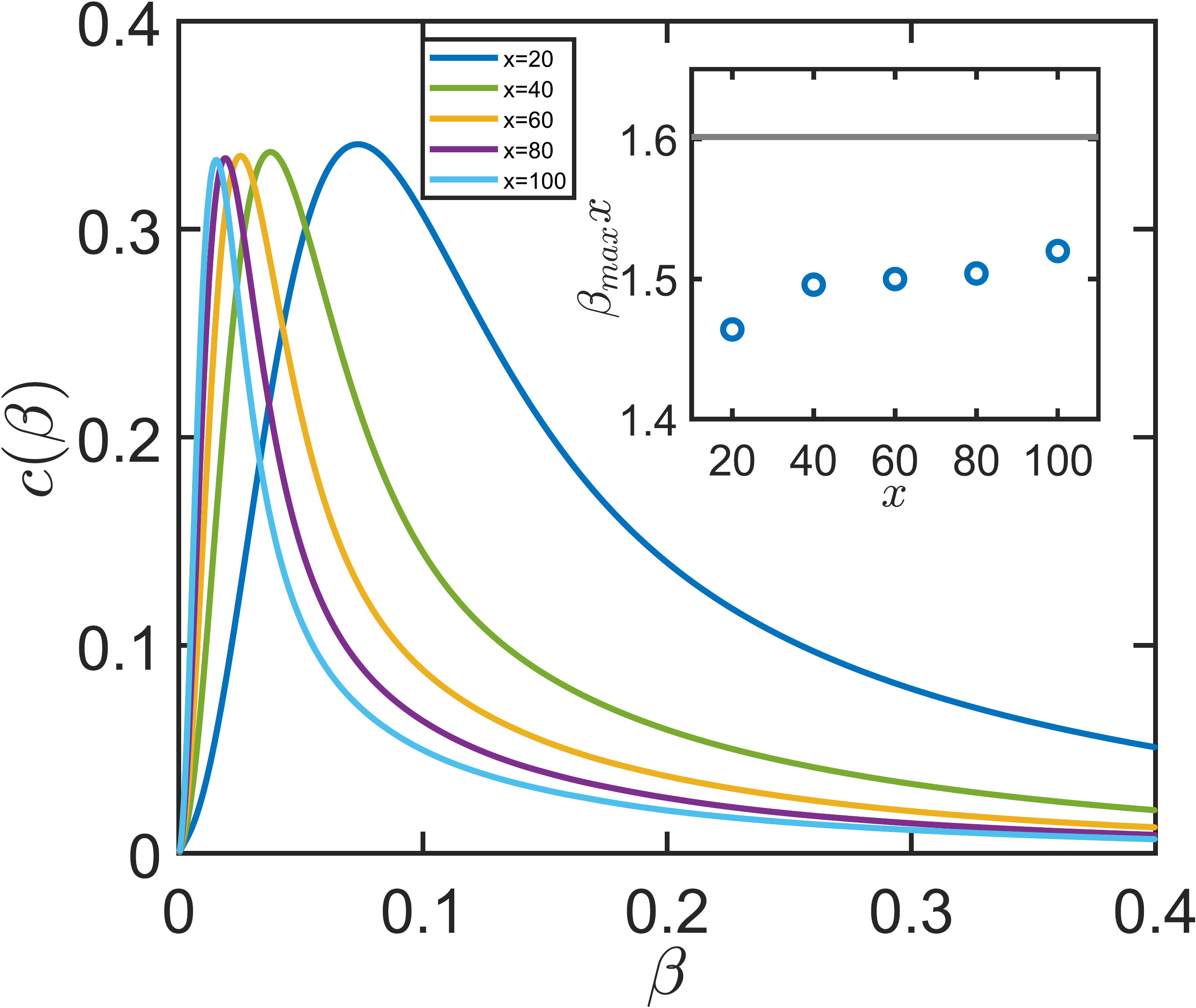}\par\medskip
\caption{The specific heat for a system size with $N=80$, for different values of $x$. As the lattice spacing decreases and $x$ increases, the peak of the specific heat becomes sharper.
The inset shows how the rescaled inverse temperature, $\beta_{\max}x$, with $\beta_{\max}$ the inverse temperature that corresponds to the peak, behaves as we increase the parameter $x$. The grey line indicates the exact value for the XY model, to which the Hamiltonian converges in the limit $x\to\infty$.
The specific heat was computed using standard MPS methods. } 
\label{The specific heat}
\end{figure}

In Fig.~\ref{fig: condensateN20N30} we plot the energy, the entropy and the condensate obtained with this method for system sizes $N=20$ and $N=30$ and $x=5$. It is also possible to compute the observables by finding a MPO approximation to the Gibbs ensemble of the model~\cite{verstraete2004,zwolak2004,schollwoeck2011}. For the Schwinger model this MPO produces very precise results for thermal observables~\cite{banuls2015thermal,banuls2016thermal,Buyens:2016ecr},  and we use it here as reference
\footnote{Even though the standard MPO method, which proceeds via imaginary time evolution, becomes less accurate as $\beta$ increases due to accumulation of truncation, it is possible to estimate this error systematically (see~\cite{banuls2015thermal}) and use sufficiently converged values as reference, as is done in Fig~\ref{fig: condensateN20N30}.}.
While we observe that the results obtained with the method of section~\ref{sec:Methods} (blue line) agree well with the reference (grey line) for the energy, the figure shows that, specially for the condensate, the agreement only holds for small values of $\beta$. 
We attribute the limitations of our calculation to the fact that for larger values of $\beta$, the Boltzmann factor $e^{-\beta E}$ in the integrals enhances the contributions of the lower edge of the spectrum, where the DOS is many orders of magnitude smaller than at the peak and the truncated approximation is less accurate. 
For more detailed explanation see Sec.~\ref{sec: convergenceErrors}. 

\begin{figure*}[t]
\centering
    \includegraphics[width=0.4\linewidth]{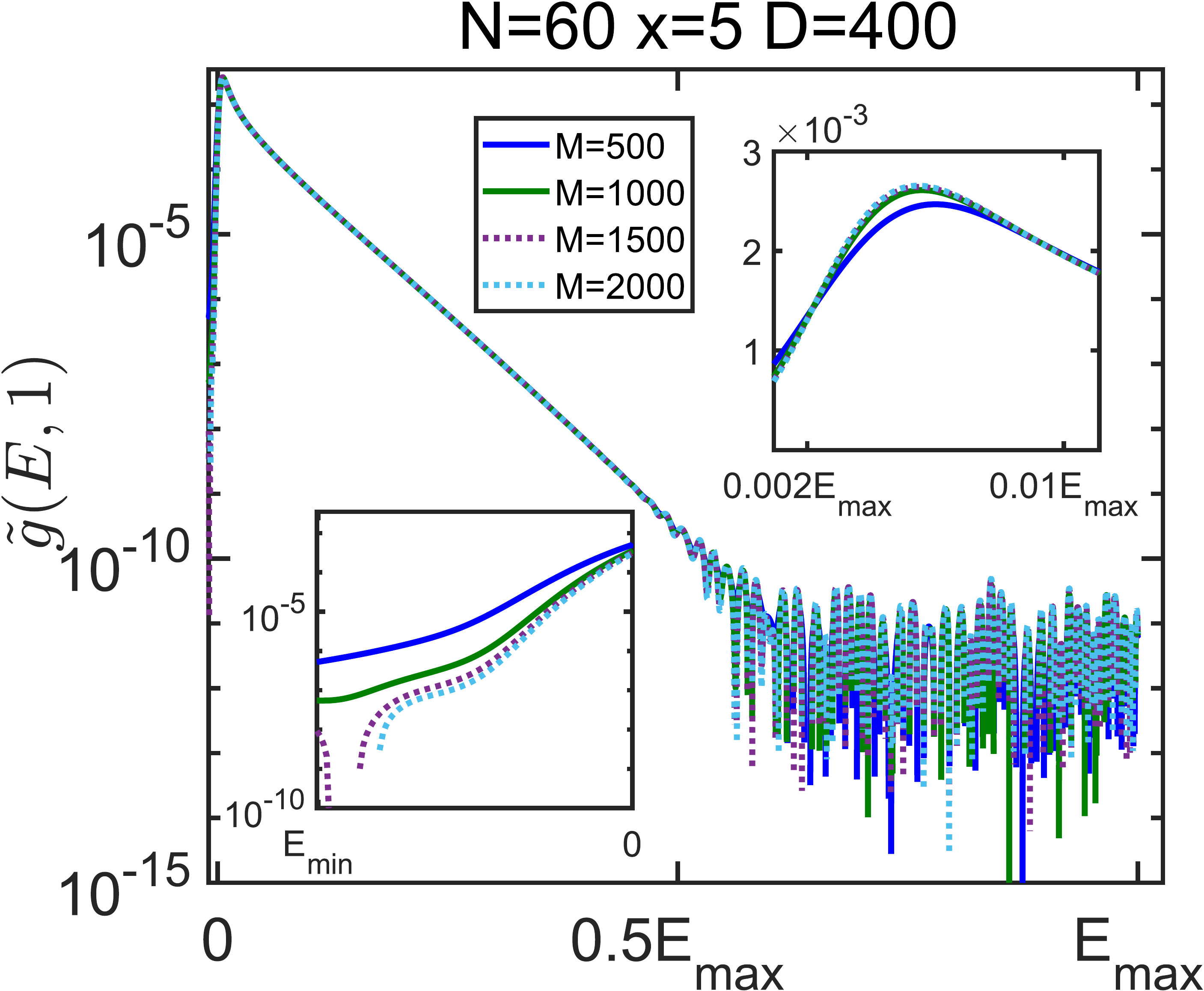}\hfil
    \includegraphics[width=0.4\linewidth]{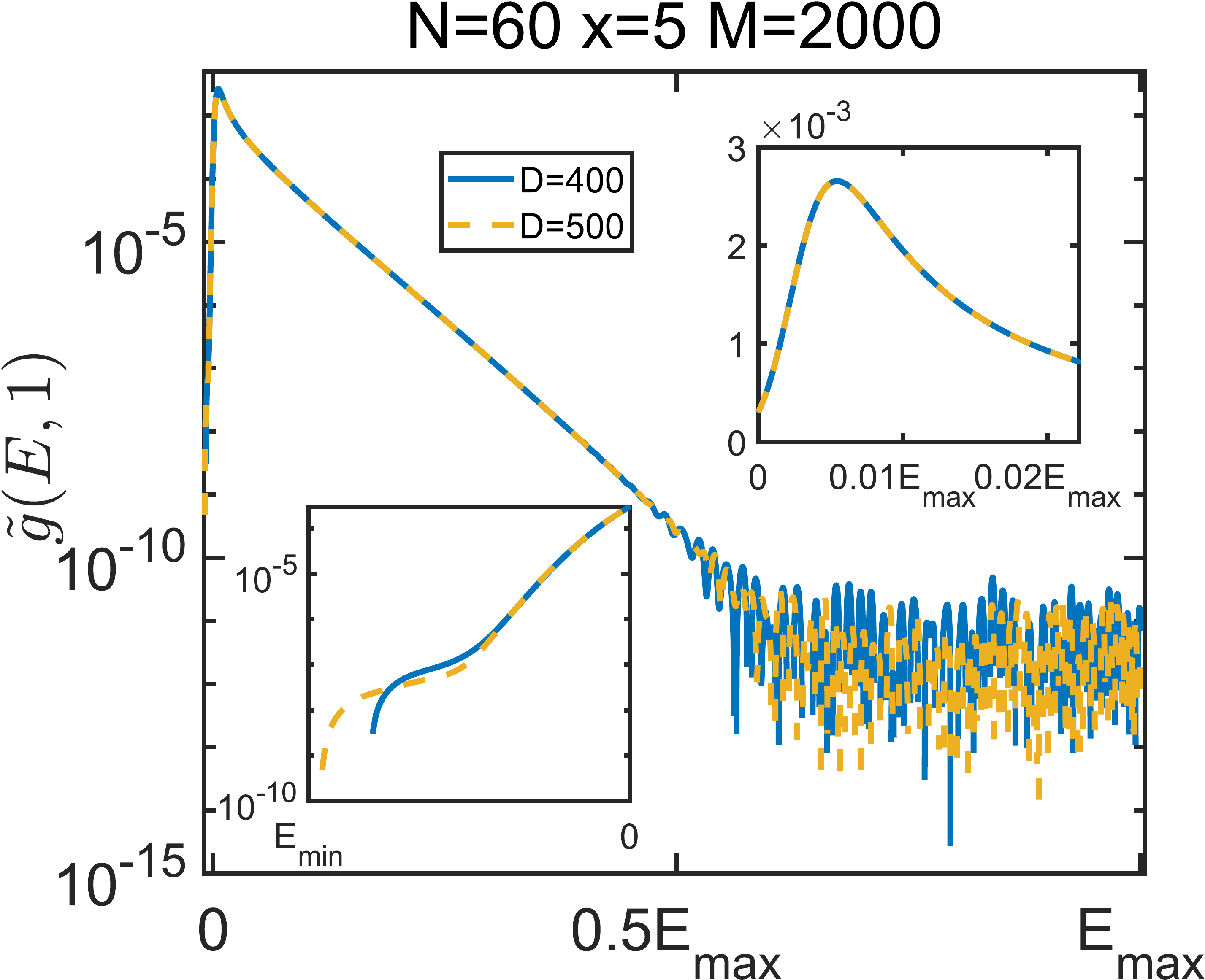}\par\medskip
    \includegraphics[width=0.4\linewidth]{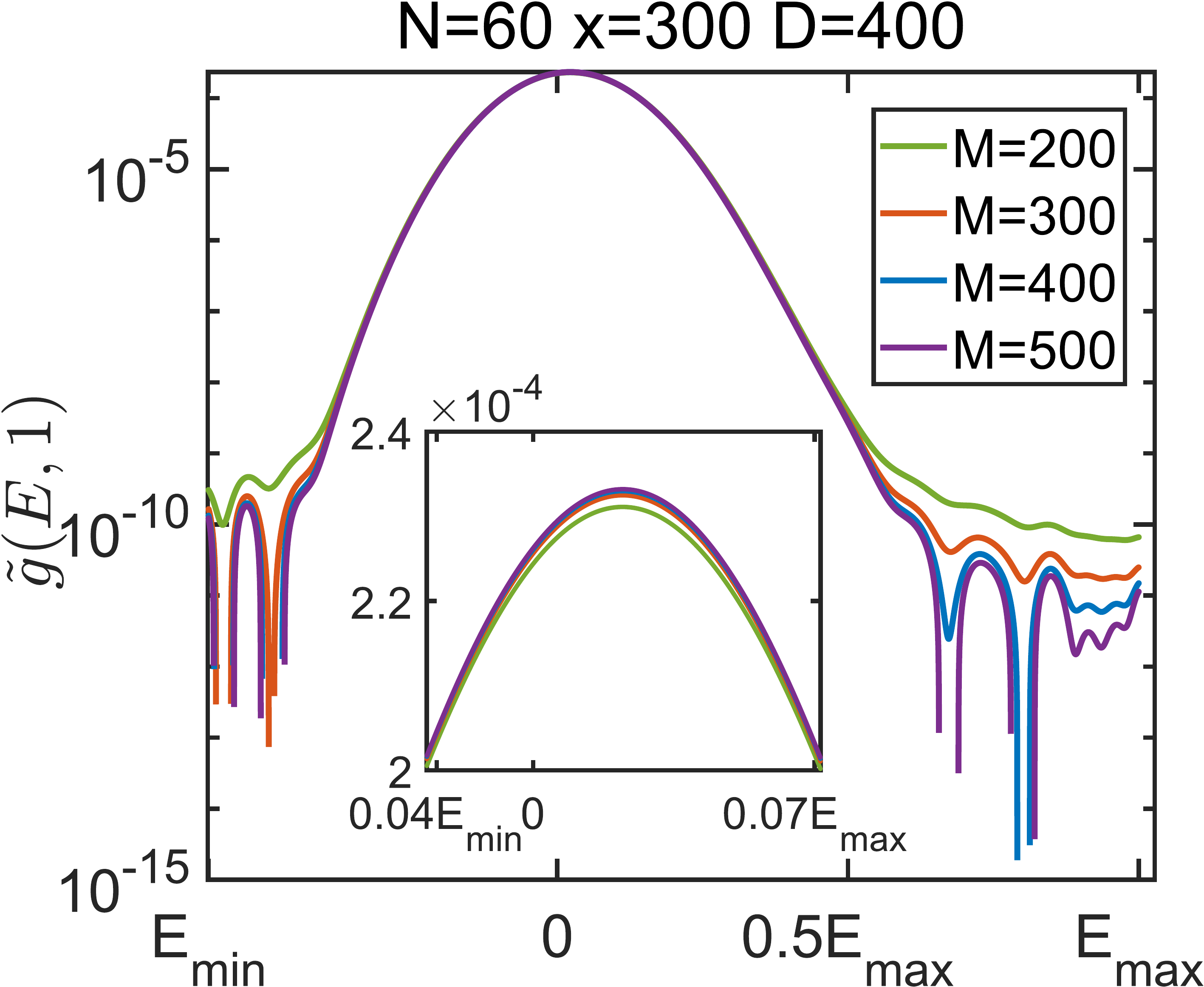}\hfil
    \includegraphics[width=0.4\linewidth]{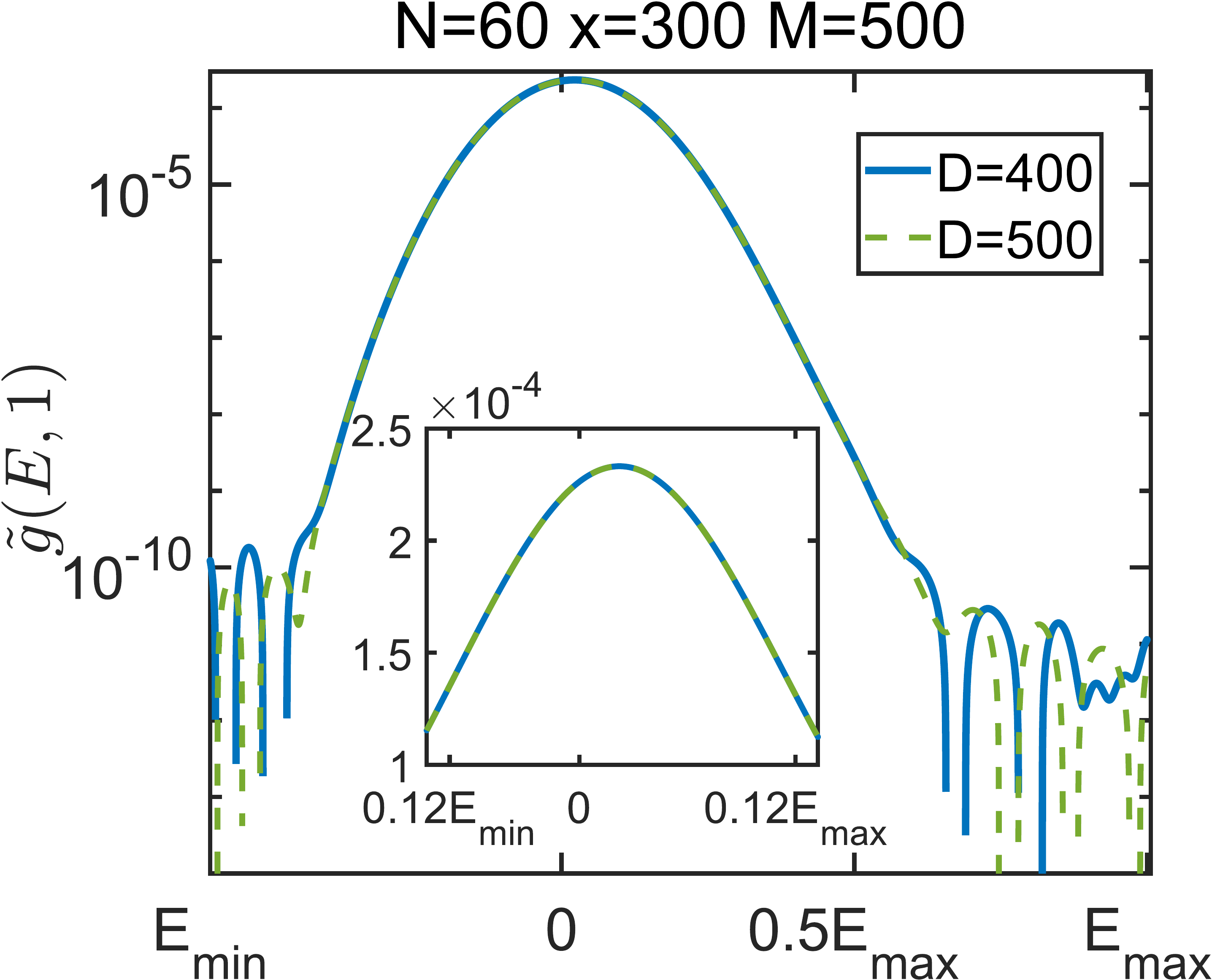}
\caption{Error analysis in the calculation of the DOS for a system of size $N=60$ and different lattice spacing $x=5$ (upper panels) and $x=300$ (lower panels). The left panels show the effect of truncating the Chebyshev series to finite value $M$, using large enough bond dimension $D$ for convergence of all the cases (as demonstrated by the right panels).
}
\label{convergence}
\end{figure*}

In~\cite{elser1996qed} it was suggested that the shape of the DOS gave rise to a peak in the specific heat Eq.~\eqref{eq: specific heat} which appeared to diverge in the continuum limit.  These conclusions  were contradicted by later, more precise studies~\cite{Strauss2008prl,Strauss2009pos}, and attributed to the limited precision of the simulations which would not allow for a reliable continuum extrapolation. It is interesting to investigate whether limiting the study to finite systems and lattice spacing could show similar spurious signals in the specific heat.
To this end, and because the precision obtained with our method is limited for large systems, we examine the specific heat using a standard MPS simulation~\cite{banuls2015thermal,banuls2016thermal,Buyens:2016ecr}.
We fix a system size $N=80$ and systematically increase the parameter $x$. 
Notice that in this way we are exploring properties of the (finite) lattice model, since this does not correspond to approaching the continuum (which would require increasing the system size as $N\propto \sqrt{x}$ to maintain a consistent physical volume). Figure~\ref{The specific heat} shows the dependence of the  specific heat with the inverse temperature as $x$ increases. We observe a smooth peak in the specific heat, that drifts slowly towards infinite temperature as $x$ increases (as mentioned before, in the limit $x\to \infty$ the Hamiltonian reduces to the exactly solvable XY model with a multiplication factor $x$). In order to connect the observed quantities to their continuum correspondence, it is important to highlight here that $\beta$ used in figures ~\ref{fig: condensateN20N30} and ~\ref{The specific heat} is adimensional since the Hamiltonian of Eq.~\eqref{eq: Schwinger spin} is also adimensional. It is related to the physical inverse temperature $\beta_{\mathrm{phys}}$ (the inverse temperature in the continuum) in the following way~\cite{banuls2015thermal}\begin{align}
    \beta&=\frac{\beta_{\mathrm{phys}}g}{2\sqrt{x}}.
\end{align}

\subsection{Convergence and error analysis} \label{sec: convergenceErrors}

The calculated quantities,  $g_{M}(E;O)$ and $g_{M}(E;\mathds{1})$, have two sources of errors, as discussed in~\cite{yang2020probing}. One type of error comes from truncating the Chebyshev expansion up to order $M$. As shown in \cite{weisse2006kernel}, when the Jackson kernel is used, and the function being approximated is continuous in the interval $[-1,1]$, the errors are of order $O(1/M)$ and consequently are reduced as we increase the Chebyshev cut-off $M$. 
The second source is the truncation of the MPOs used to represent the polynomials to a finite bond dimension $T_{n}^{D}(\Tilde{H})$, with $D\leq D_{\max}$. The observation in~\cite{yang2020probing} indicates that the bond dimension required to approximate the $n-$th polynomial with fixed precision grows fast (even exponentially) with the order $n$.

The effect of these errors is that of effectively limiting the energy resolution attainable with a given bond dimension, since the latter limits the number of moments that can be reliably extracted. 
When applied to the DOS and related functions, written as sums of Dirac delta terms, we expect that the effect of this limited resolution becomes more significant wherever the DOS exhibits sharp features or where its absolute value is small. In our case, this corresponds to the edges of the spectrum, where eigenstates become more sparse, while the effect is minor where the DOS is sufficiently large, such as near the peak. 

In the computation of the spectral quantities $g_{M}(E;O)$ and $g_{M}(E;\mathds{1})$, only traces of the polynomials or (of their product with a given operator) appear. These appear more converged in bond dimension than the global error analyzed in~\cite{yang2020probing}, which allows us to explore relatively large values of the Chebyshev truncation parameter $M$.
In particular, we found that characterizing the 
overall shape of the DOS, as discussed in \ref{subsec:dos}, 
was possible with moderate computational resources.  Concretely, for $N=60$, we found $M=500$ to be sufficient for large values of $x$,
$100\leq x\leq300$.
For smaller $x$, the peak of the DOS is sharper and requires a larger number of polynomials, growing to $M=1000$ for $30\leq x \leq 50$ and up to $M=2000$ for $x=5$, as shown in figure~\ref{convergence}.
For all these cases, we found bond dimension $D=400$ to be enough to accurately capture the shape of the DOS (see right panels in figure~\ref{convergence}), and to have convergence of the results shown in figures~\ref{fitting} and~\ref{Area difference fitting}.

For the same parameters, however, the edges of the spectrum are not converged, as can be appreciated in Fig.~\ref{convergence}.
Although the same moments are used to compute the DOS at all energies, the final result is the sum of a large number of terms with energy dependent coefficients that can oscillate with large frequencies (increasing with the order). Near the peak the sum is large, and the relative contribution of large order moments (more affected by truncation) is small. However, close to the edges precise cancellations of terms are required to recover the much smaller value of the DOS. Hence, in such areas, the result is much more sensitive to truncation errors in the moments of all orders.
In practice, the computed DOS 
exhibits large fluctuations that depend on the truncation parameters in the regions where the true magnitude of the spectral density is small, up to the spectral edges.
Because of the small magnitude of the DOS, these fluctuations do not appear to be present in Fig.~\ref{fitting}, but they are visible when we plot the DOS in a logarithmic scale in Fig.~\ref{convergence}.

Very close to the edges, we may expect that the discreteness of the  energy spectrum for finite systems further contributes to the fluctuations.
The typical energy gaps are exponentially small in the system size, but close to the edges, the gaps can be much larger. 
In order to be able to capture the sparse spectrum in these regions with our method, $M$ would need to take much larger values, which in turn would require even bigger bond dimension $D$.

The discussion above refers to the determination of the DOS. The thermodynamic observables shown in Fig.~\ref{fig: condensateN20N30} are computed using the same approximation to the Chebyshev moments, but combined with different coefficients, to compute Eq.~\eqref{canonical ensemble}.
The truncation parameters that allow us to determine the form of the DOS are not enough to explore the whole range of temperatures.
Instead, even if the errors at the edges do not affect the form of the DOS, they do constrain the interval of temperatures that is accessible when calculating thermodynamic quantities with our method (see Fig.~\ref{fig: condensateN20N30}). Specifically, as $\beta$ increases, because of the Boltzmann factor, so does the contribution from the energies near the lowest edge of the spectrum $E_{min}$ to the quantities $g(E;\mathds{1})$ and $g(E;O)$. 
 Because the computation with fixed truncation parameters is less accurate near the edges,  attaining a precise calculation of the different thermodynamic quantities becomes more computationally expensive as $\beta$ increases. The truncation errors related to the bond dimension $D$ at the edges of the spectrum appear to be more significant as we increase the system size $N$, consistent with a narrower spectral density with much smaller relative magnitude of the DOS in the tails of the distribution, as compared to the peak, and smaller gaps that require better energy resolution.
 We do not carry out a detailed error analysis for the observables, but instead show in Fig.~\ref{fig: condensateN20N30} the best results obtained, as compared to the exact results. These correspond to the biggest value of $M$ that is accessible and a fixed bond dimension $D=500$.
We notice that, with the given bond dimension, our results seem to improve significantly as we increase $M$, for most of the observables of interest.

\section{Discussion}\label{sec:discussion}

In this work, we have applied the recently introduced method~\cite{yang2020probing} to characterize the density of states of the lattice Schwinger model.This technique gives access to the overall shape of the DOS over the full range of energies and can be used to derive thermodynamic quantities that are expressed as integrals of the DOS with Boltzmann factors. 

We have shown how the shape of the distribution for the lattice Schwinger model changes with the system size and the lattice spacing parameter, from a very asymmetric spectrum at large spacings, to an approximately Gaussian form for 
sufficiently small ones. It is worth noticing that, for any finite system size, there appear to be values of the parameters for which the DOS strongly deviates from a Gaussian behavior.
Furthermore, the density of states for  large lattice spacing exhibits a sharper feature at low energies, with a very fast increase, and a narrow peak, decaying much more slowly towards high energies.

Our observations on the spectral properties may be relevant for the interpretation of experimental realizations of the model, which has already been simulated in some pioneering quantum simulations.
Since experiments are necessarily limited to finite (for the moment relatively small) system sizes, it is interesting to understand which ranges of
 parameters ensure a behavior close to the continuum limit for the spectrum, and the thermal properties.
 Our analysis can be similarly carried out for other LGT models in one spatial dimension.

Regarding physical observables, our method gives in principle access to all thermodynamic quantities. 
However, we found that in practice the computation is limited to only relatively high temperatures, and exploring lower temperatures has a much higher computational cost compared to the determination of the DOS. 

In conclusion, the method employed here opens another aspect of the quantum many-body problem, and in particular LGT, to the exploration with tensor networks.
The DOS method allows us to distinguish the features of the density of states, but directly using it
for thermodynamic observables seems to be limited to relatively high temperatures. 
Since this limitation does not affect other standard MPS techniques, a combination of methods seems to be the most promising strategy to explore other LGT problems in the future.

\appendix

\section{}\label{Appendix A}

For the sake of completeness, we summarize here the basic ingredients of the expansions used in the paper. A complete review can be found in ~\cite{weisse2006kernel}.

In general, any function $f(x)$ that is piecewise and continuous, with $x$ within the interval $[-1,1]$, can be written as an  expansion in terms of Chebyshev polynomials,
 \begin{align}~\label{Chebyshev expansion infinite}
f(x) = \frac{1}{\pi \sqrt{1-x^{2}}}\left[\mu_{0}+2\sum_{n=1}^{\infty}\mu_{n}T_{n}(x)\right],
\end{align} 
where $T_{n}(x)$ are the Chebyshev polynomials of the first kind, defined for non-negative order $n$, and the coefficients of the expansion are determined by the moments $\mu_n=\int_{-1}^1 f(x) T_n(x) dx$.
The polynomials can be defined by their recurrence relation
\begin{align}
T_{n+2}(x) = 2xT_{n+1}(x)-T_{n}(x), 
\label{Chebyshev pol}
\end{align}
with the first two polynomials
polynomials defined as $T_{0}(x)=1$ and $T_{1}(x)=x$.
Furthermore, the polynomials $T_{n}(x)$ satisfy the following orthogonality relations   
\begin{align}
\langle T_{n}|T_{m} \rangle = \int_{-1}^{1} \frac{T_{n}(x)T_{m}(x)}{\sqrt{1-x^{2}}}dx = (1+\delta_{n,0})\frac{\pi}{2} \delta_{n,m},
\label{Chebyshev orthogonality}
\end{align}
Truncating the sum of Eq.~\eqref{Chebyshev expansion infinite} to a finite order $M$ defines an approximation to the function.
The kernel polynomial method (KPM) improves the properties of the approximation by including additional coefficients $\gamma_{n}^{M}$, which can be chosen in order to suppress the oscillations that come from having a finite series of  polynomials~\cite{weisse2006kernel}.
The corresponding approximation 
$f_{M}(x)$, for $M$ terms in the sum reads
\begin{align}\label{Chebyshev expansion}
f_{M}(x) := \frac{1}{\pi \sqrt{1-x^{2}}}\left[\gamma_{0}^{M}\mu_{0}+2\sum_{n=1}^{M-1}\gamma_{n}^{M}\mu_{n}T_{n}(x)\right].
\end{align}
 
For the Jackson kernel, which we use in our calculations, 
\begin{align}
\gamma_{n}^{M} = \frac{(M-n+1)\cos\frac{\pi n}{M+1}+\sin\frac{\pi n}{M+1}\cot\frac{\pi}{M+1}}{M+1}.
\label{Jackson}
\end{align}

\begin{acknowledgments}
We thank Sofyan Iblisdir and Yilun Yang for very helpful discussions. This work was
partly supported by the Deutsche Forschungsgemeinschaft
(DFG, German Research Foundation) under Germany’s
Excellence Strategy—EXC-2111—390814868, and EU-QUANTERA project QTFLAG (BMBF Grant
No. 13N14780).
\end{acknowledgments}


\begin{thebibliography}{56}%
\makeatletter
\providecommand \@ifxundefined [1]{%
 \@ifx{#1\undefined}
}%
\providecommand \@ifnum [1]{%
 \ifnum #1\expandafter \@firstoftwo
 \else \expandafter \@secondoftwo
 \fi
}%
\providecommand \@ifx [1]{%
 \ifx #1\expandafter \@firstoftwo
 \else \expandafter \@secondoftwo
 \fi
}%
\providecommand \natexlab [1]{#1}%
\providecommand \enquote  [1]{``#1''}%
\providecommand \bibnamefont  [1]{#1}%
\providecommand \bibfnamefont [1]{#1}%
\providecommand \citenamefont [1]{#1}%
\providecommand \href@noop [0]{\@secondoftwo}%
\providecommand \href [0]{\begingroup \@sanitize@url \@href}%
\providecommand \@href[1]{\@@startlink{#1}\@@href}%
\providecommand \@@href[1]{\endgroup#1\@@endlink}%
\providecommand \@sanitize@url [0]{\catcode `\\12\catcode `\$12\catcode
  `\&12\catcode `\#12\catcode `\^12\catcode `\_12\catcode `\%12\relax}%
\providecommand \@@startlink[1]{}%
\providecommand \@@endlink[0]{}%
\providecommand \url  [0]{\begingroup\@sanitize@url \@url }%
\providecommand \@url [1]{\endgroup\@href {#1}{\urlprefix }}%
\providecommand \urlprefix  [0]{URL }%
\providecommand \Eprint [0]{\href }%
\providecommand \doibase [0]{https://doi.org/}%
\providecommand \selectlanguage [0]{\@gobble}%
\providecommand \bibinfo  [0]{\@secondoftwo}%
\providecommand \bibfield  [0]{\@secondoftwo}%
\providecommand \translation [1]{[#1]}%
\providecommand \BibitemOpen [0]{}%
\providecommand \bibitemStop [0]{}%
\providecommand \bibitemNoStop [0]{.\EOS\space}%
\providecommand \EOS [0]{\spacefactor3000\relax}%
\providecommand \BibitemShut  [1]{\csname bibitem#1\endcsname}%
\let\auto@bib@innerbib\@empty
\bibitem [{\citenamefont {Metropolis}\ \emph {et~al.}(1953)\citenamefont
  {Metropolis}, \citenamefont {Rosenbluth}, \citenamefont {Rosenbluth},
  \citenamefont {Teller},\ and\ \citenamefont {Teller}}]{Metropolis1953}%
  \BibitemOpen
  \bibfield  {author} {\bibinfo {author} {\bibfnamefont {N.}~\bibnamefont
  {Metropolis}}, \bibinfo {author} {\bibfnamefont {A.~W.}\ \bibnamefont
  {Rosenbluth}}, \bibinfo {author} {\bibfnamefont {M.~N.}\ \bibnamefont
  {Rosenbluth}}, \bibinfo {author} {\bibfnamefont {A.~H.}\ \bibnamefont
  {Teller}},\ and\ \bibinfo {author} {\bibfnamefont {E.}~\bibnamefont
  {Teller}},\ }\href {https://doi.org/10.1063/1.1699114} {\bibfield  {journal}
  {\bibinfo  {journal} {The Journal of Chemical Physics}\ }\textbf {\bibinfo
  {volume} {21}},\ \bibinfo {pages} {1087} (\bibinfo {year}
  {1953})}\BibitemShut {NoStop}%
\bibitem [{\citenamefont {{Schollw{\"o}ck}}(2011)}]{schollwoeck2011}%
  \BibitemOpen
  \bibfield  {author} {\bibinfo {author} {\bibfnamefont {U.}~\bibnamefont
  {{Schollw{\"o}ck}}},\ }\href {https://doi.org/10.1016/j.aop.2010.09.012}
  {\bibfield  {journal} {\bibinfo  {journal} {Annals of Physics}\ }\textbf
  {\bibinfo {volume} {326}},\ \bibinfo {pages} {96} (\bibinfo {year} {2011})},\
  \Eprint {https://arxiv.org/abs/1008.3477} {arXiv:1008.3477 [cond-mat.str-el]}
  \BibitemShut {NoStop}%
\bibitem [{\citenamefont {{Verstraete}}\ \emph {et~al.}(2008)\citenamefont
  {{Verstraete}}, \citenamefont {{Murg}},\ and\ \citenamefont
  {{Cirac}}}]{verstraete2008}%
  \BibitemOpen
  \bibfield  {author} {\bibinfo {author} {\bibfnamefont {F.}~\bibnamefont
  {{Verstraete}}}, \bibinfo {author} {\bibfnamefont {V.}~\bibnamefont
  {{Murg}}},\ and\ \bibinfo {author} {\bibfnamefont {J.~I.}\ \bibnamefont
  {{Cirac}}},\ }\href {https://doi.org/10.1080/14789940801912366} {\bibfield
  {journal} {\bibinfo  {journal} {Advances in Physics}\ }\textbf {\bibinfo
  {volume} {57}},\ \bibinfo {pages} {143} (\bibinfo {year} {2008})},\ \Eprint
  {https://arxiv.org/abs/0907.2796} {arXiv:0907.2796 [quant-ph]} \BibitemShut
  {NoStop}%
\bibitem [{\citenamefont {Orus}(2014)}]{orus2014}%
  \BibitemOpen
  \bibfield  {author} {\bibinfo {author} {\bibfnamefont {R.}~\bibnamefont
  {Orus}},\ }\href {https://doi.org/10.1016/j.aop.2014.06.013} {\bibfield
  {journal} {\bibinfo  {journal} {Annals Phys.}\ }\textbf {\bibinfo {volume}
  {349}},\ \bibinfo {pages} {117} (\bibinfo {year} {2014})},\ \Eprint
  {https://arxiv.org/abs/1306.2164} {arXiv:1306.2164 [cond-mat.str-el]}
  \BibitemShut {NoStop}%
\bibitem [{\citenamefont {Langfeld}\ \emph {et~al.}(2012)\citenamefont
  {Langfeld}, \citenamefont {Lucini},\ and\ \citenamefont
  {Rago}}]{langfeld2012density}%
  \BibitemOpen
  \bibfield  {author} {\bibinfo {author} {\bibfnamefont {K.}~\bibnamefont
  {Langfeld}}, \bibinfo {author} {\bibfnamefont {B.}~\bibnamefont {Lucini}},\
  and\ \bibinfo {author} {\bibfnamefont {A.}~\bibnamefont {Rago}},\ }\href
  {https://doi.org/10.1103/PhysRevLett.109.111601} {\bibfield  {journal}
  {\bibinfo  {journal} {Phys. Rev. Lett.}\ }\textbf {\bibinfo {volume} {109}},\
  \bibinfo {pages} {111601} (\bibinfo {year} {2012})},\ \Eprint
  {https://arxiv.org/abs/1204.3243} {arXiv:1204.3243 [hep-lat]} \BibitemShut
  {NoStop}%
\bibitem [{\citenamefont {Troyer}\ and\ \citenamefont
  {Wiese}(2005)}]{Troyer2005}%
  \BibitemOpen
  \bibfield  {author} {\bibinfo {author} {\bibfnamefont {M.}~\bibnamefont
  {Troyer}}\ and\ \bibinfo {author} {\bibfnamefont {U.-J.}\ \bibnamefont
  {Wiese}},\ }\href {https://doi.org/10.1103/PhysRevLett.94.170201} {\bibfield
  {journal} {\bibinfo  {journal} {Phys. Rev. Lett.}\ }\textbf {\bibinfo
  {volume} {94}},\ \bibinfo {pages} {170201} (\bibinfo {year}
  {2005})}\BibitemShut {NoStop}%
\bibitem [{\citenamefont {Wang}\ and\ \citenamefont {Landau}(2001)}]{Wang2001}%
  \BibitemOpen
  \bibfield  {author} {\bibinfo {author} {\bibfnamefont {F.}~\bibnamefont
  {Wang}}\ and\ \bibinfo {author} {\bibfnamefont {D.~P.}\ \bibnamefont
  {Landau}},\ }\href {https://doi.org/10.1103/PhysRevLett.86.2050} {\bibfield
  {journal} {\bibinfo  {journal} {Phys. Rev. Lett.}\ }\textbf {\bibinfo
  {volume} {86}},\ \bibinfo {pages} {2050} (\bibinfo {year}
  {2001})}\BibitemShut {NoStop}%
\bibitem [{\citenamefont {Trebst}\ \emph {et~al.}(2004)\citenamefont {Trebst},
  \citenamefont {Huse},\ and\ \citenamefont {Troyer}}]{Trebst2004}%
  \BibitemOpen
  \bibfield  {author} {\bibinfo {author} {\bibfnamefont {S.}~\bibnamefont
  {Trebst}}, \bibinfo {author} {\bibfnamefont {D.~A.}\ \bibnamefont {Huse}},\
  and\ \bibinfo {author} {\bibfnamefont {M.}~\bibnamefont {Troyer}},\ }\href
  {https://doi.org/10.1103/PhysRevE.70.046701} {\bibfield  {journal} {\bibinfo
  {journal} {Phys. Rev. E}\ }\textbf {\bibinfo {volume} {70}},\ \bibinfo
  {pages} {046701} (\bibinfo {year} {2004})}\BibitemShut {NoStop}%
\bibitem [{\citenamefont {Troyer}\ \emph {et~al.}(2003)\citenamefont {Troyer},
  \citenamefont {Wessel},\ and\ \citenamefont {Alet}}]{Troyer2003}%
  \BibitemOpen
  \bibfield  {author} {\bibinfo {author} {\bibfnamefont {M.}~\bibnamefont
  {Troyer}}, \bibinfo {author} {\bibfnamefont {S.}~\bibnamefont {Wessel}},\
  and\ \bibinfo {author} {\bibfnamefont {F.}~\bibnamefont {Alet}},\ }\href
  {https://doi.org/10.1103/PhysRevLett.90.120201} {\bibfield  {journal}
  {\bibinfo  {journal} {Phys. Rev. Lett.}\ }\textbf {\bibinfo {volume} {90}},\
  \bibinfo {pages} {120201} (\bibinfo {year} {2003})}\BibitemShut {NoStop}%
\bibitem [{\citenamefont {Wessel}\ \emph {et~al.}(2007)\citenamefont {Wessel},
  \citenamefont {Stoop}, \citenamefont {Gull}, \citenamefont {Trebst},\ and\
  \citenamefont {Troyer}}]{Wessel2007}%
  \BibitemOpen
  \bibfield  {author} {\bibinfo {author} {\bibfnamefont {S.}~\bibnamefont
  {Wessel}}, \bibinfo {author} {\bibfnamefont {N.}~\bibnamefont {Stoop}},
  \bibinfo {author} {\bibfnamefont {E.}~\bibnamefont {Gull}}, \bibinfo {author}
  {\bibfnamefont {S.}~\bibnamefont {Trebst}},\ and\ \bibinfo {author}
  {\bibfnamefont {M.}~\bibnamefont {Troyer}},\ }\href
  {https://doi.org/10.1088/1742-5468/2007/12/p12005} {\bibfield  {journal}
  {\bibinfo  {journal} {Journal of Statistical Mechanics: Theory and
  Experiment}\ }\textbf {\bibinfo {volume} {2007}},\ \bibinfo {pages} {P12005}
  (\bibinfo {year} {2007})}\BibitemShut {NoStop}%
\bibitem [{\citenamefont {Wei\ss{}e}\ \emph {et~al.}(2006)\citenamefont
  {Wei\ss{}e}, \citenamefont {Wellein}, \citenamefont {Alvermann},\ and\
  \citenamefont {Fehske}}]{weisse2006kernel}%
  \BibitemOpen
  \bibfield  {author} {\bibinfo {author} {\bibfnamefont {A.}~\bibnamefont
  {Wei\ss{}e}}, \bibinfo {author} {\bibfnamefont {G.}~\bibnamefont {Wellein}},
  \bibinfo {author} {\bibfnamefont {A.}~\bibnamefont {Alvermann}},\ and\
  \bibinfo {author} {\bibfnamefont {H.}~\bibnamefont {Fehske}},\ }\href
  {https://doi.org/10.1103/RevModPhys.78.275} {\bibfield  {journal} {\bibinfo
  {journal} {Rev. Mod. Phys.}\ }\textbf {\bibinfo {volume} {78}},\ \bibinfo
  {pages} {275} (\bibinfo {year} {2006})}\BibitemShut {NoStop}%
\bibitem [{\citenamefont {Schrodi}\ \emph {et~al.}(2017)\citenamefont
  {Schrodi}, \citenamefont {Silvi}, \citenamefont {Tschirsich}, \citenamefont
  {Fazio},\ and\ \citenamefont {Montangero}}]{schrodi2017density}%
  \BibitemOpen
  \bibfield  {author} {\bibinfo {author} {\bibfnamefont {F.}~\bibnamefont
  {Schrodi}}, \bibinfo {author} {\bibfnamefont {P.}~\bibnamefont {Silvi}},
  \bibinfo {author} {\bibfnamefont {F.}~\bibnamefont {Tschirsich}}, \bibinfo
  {author} {\bibfnamefont {R.}~\bibnamefont {Fazio}},\ and\ \bibinfo {author}
  {\bibfnamefont {S.}~\bibnamefont {Montangero}},\ }\href
  {https://doi.org/10.1103/PhysRevB.96.094303} {\bibfield  {journal} {\bibinfo
  {journal} {Phys. Rev. B}\ }\textbf {\bibinfo {volume} {96}},\ \bibinfo
  {pages} {094303} (\bibinfo {year} {2017})}\BibitemShut {NoStop}%
\bibitem [{\citenamefont {Yang}\ \emph
  {et~al.}(2020{\natexlab{a}})\citenamefont {Yang}, \citenamefont {Iblisdir},
  \citenamefont {Cirac},\ and\ \citenamefont {Ba\~nuls}}]{yang2020probing}%
  \BibitemOpen
  \bibfield  {author} {\bibinfo {author} {\bibfnamefont {Y.}~\bibnamefont
  {Yang}}, \bibinfo {author} {\bibfnamefont {S.}~\bibnamefont {Iblisdir}},
  \bibinfo {author} {\bibfnamefont {J.~I.}\ \bibnamefont {Cirac}},\ and\
  \bibinfo {author} {\bibfnamefont {M.~C.}\ \bibnamefont {Ba\~nuls}},\ }\href
  {https://doi.org/10.1103/PhysRevLett.124.100602} {\bibfield  {journal}
  {\bibinfo  {journal} {Phys. Rev. Lett.}\ }\textbf {\bibinfo {volume} {124}},\
  \bibinfo {pages} {100602} (\bibinfo {year} {2020}{\natexlab{a}})}\BibitemShut
  {NoStop}%
\bibitem [{\citenamefont {Verstraete}\ \emph
  {et~al.}(2004{\natexlab{a}})\citenamefont {Verstraete}, \citenamefont
  {Garc\'{\i}a-Ripoll},\ and\ \citenamefont {Cirac}}]{verstraete2004}%
  \BibitemOpen
  \bibfield  {author} {\bibinfo {author} {\bibfnamefont {F.}~\bibnamefont
  {Verstraete}}, \bibinfo {author} {\bibfnamefont {J.~J.}\ \bibnamefont
  {Garc\'{\i}a-Ripoll}},\ and\ \bibinfo {author} {\bibfnamefont {J.~I.}\
  \bibnamefont {Cirac}},\ }\href
  {https://doi.org/10.1103/PhysRevLett.93.207204} {\bibfield  {journal}
  {\bibinfo  {journal} {Phys. Rev. Lett.}\ }\textbf {\bibinfo {volume} {93}},\
  \bibinfo {pages} {207204} (\bibinfo {year} {2004}{\natexlab{a}})}\BibitemShut
  {NoStop}%
\bibitem [{\citenamefont {Pirvu}\ \emph {et~al.}(2010)\citenamefont {Pirvu},
  \citenamefont {Murg}, \citenamefont {Cirac},\ and\ \citenamefont
  {Verstraete}}]{pirvu2010}%
  \BibitemOpen
  \bibfield  {author} {\bibinfo {author} {\bibfnamefont {B.}~\bibnamefont
  {Pirvu}}, \bibinfo {author} {\bibfnamefont {V.}~\bibnamefont {Murg}},
  \bibinfo {author} {\bibfnamefont {J.~I.}\ \bibnamefont {Cirac}},\ and\
  \bibinfo {author} {\bibfnamefont {F.}~\bibnamefont {Verstraete}},\ }\href
  {https://doi.org/10.1088/1367-2630/12/2/025012} {\bibfield  {journal}
  {\bibinfo  {journal} {New Journal of Physics}\ }\textbf {\bibinfo {volume}
  {12}},\ \bibinfo {pages} {025012} (\bibinfo {year} {2010})}\BibitemShut
  {NoStop}%
\bibitem [{\citenamefont {Zwolak}\ and\ \citenamefont
  {Vidal}(2004)}]{zwolak2004}%
  \BibitemOpen
  \bibfield  {author} {\bibinfo {author} {\bibfnamefont {M.}~\bibnamefont
  {Zwolak}}\ and\ \bibinfo {author} {\bibfnamefont {G.}~\bibnamefont {Vidal}},\
  }\href {https://doi.org/10.1103/PhysRevLett.93.207205} {\bibfield  {journal}
  {\bibinfo  {journal} {Phys. Rev. Lett.}\ }\textbf {\bibinfo {volume} {93}},\
  \bibinfo {pages} {207205} (\bibinfo {year} {2004})}\BibitemShut {NoStop}%
\bibitem [{\citenamefont {Langfeld}\ \emph {et~al.}(2016)\citenamefont
  {Langfeld}, \citenamefont {Lucini}, \citenamefont {Pellegrini},\ and\
  \citenamefont {Rago}}]{langfeld2016efficient}%
  \BibitemOpen
  \bibfield  {author} {\bibinfo {author} {\bibfnamefont {K.}~\bibnamefont
  {Langfeld}}, \bibinfo {author} {\bibfnamefont {B.}~\bibnamefont {Lucini}},
  \bibinfo {author} {\bibfnamefont {R.}~\bibnamefont {Pellegrini}},\ and\
  \bibinfo {author} {\bibfnamefont {A.}~\bibnamefont {Rago}},\ }\href
  {https://doi.org/10.1140/epjc/s10052-016-4142-5} {\bibfield  {journal}
  {\bibinfo  {journal} {Eur. Phys. J. C}\ }\textbf {\bibinfo {volume} {76}},\
  \bibinfo {pages} {306} (\bibinfo {year} {2016})},\ \Eprint
  {https://arxiv.org/abs/1509.08391} {arXiv:1509.08391 [hep-lat]} \BibitemShut
  {NoStop}%
\bibitem [{\citenamefont {Gattringer}\ and\ \citenamefont
  {T\"orek}(2015)}]{gattringer2015density}%
  \BibitemOpen
  \bibfield  {author} {\bibinfo {author} {\bibfnamefont {C.}~\bibnamefont
  {Gattringer}}\ and\ \bibinfo {author} {\bibfnamefont {P.}~\bibnamefont
  {T\"orek}},\ }\href {https://doi.org/10.1016/j.physletb.2015.06.017}
  {\bibfield  {journal} {\bibinfo  {journal} {Phys. Lett. B}\ }\textbf
  {\bibinfo {volume} {747}},\ \bibinfo {pages} {545} (\bibinfo {year}
  {2015})},\ \Eprint {https://arxiv.org/abs/1503.04947} {arXiv:1503.04947
  [hep-lat]} \BibitemShut {NoStop}%
\bibitem [{\citenamefont {Giuliani}\ \emph {et~al.}(2016)\citenamefont
  {Giuliani}, \citenamefont {Gattringer},\ and\ \citenamefont
  {T\"orek}}]{giuliani2016developing}%
  \BibitemOpen
  \bibfield  {author} {\bibinfo {author} {\bibfnamefont {M.}~\bibnamefont
  {Giuliani}}, \bibinfo {author} {\bibfnamefont {C.}~\bibnamefont
  {Gattringer}},\ and\ \bibinfo {author} {\bibfnamefont {P.}~\bibnamefont
  {T\"orek}},\ }\href {https://doi.org/10.1016/j.nuclphysb.2016.10.005}
  {\bibfield  {journal} {\bibinfo  {journal} {Nucl. Phys. B}\ }\textbf
  {\bibinfo {volume} {913}},\ \bibinfo {pages} {627} (\bibinfo {year}
  {2016})},\ \Eprint {https://arxiv.org/abs/1607.07340} {arXiv:1607.07340
  [hep-lat]} \BibitemShut {NoStop}%
\bibitem [{\citenamefont {Giuliani}\ and\ \citenamefont
  {Gattringer}(2017)}]{giuliani2017density}%
  \BibitemOpen
  \bibfield  {author} {\bibinfo {author} {\bibfnamefont {M.}~\bibnamefont
  {Giuliani}}\ and\ \bibinfo {author} {\bibfnamefont {C.}~\bibnamefont
  {Gattringer}},\ }\href {https://doi.org/10.1016/j.physletb.2017.08.014}
  {\bibfield  {journal} {\bibinfo  {journal} {Phys. Lett. B}\ }\textbf
  {\bibinfo {volume} {773}},\ \bibinfo {pages} {166} (\bibinfo {year}
  {2017})},\ \Eprint {https://arxiv.org/abs/1703.03614} {arXiv:1703.03614
  [hep-lat]} \BibitemShut {NoStop}%
\bibitem [{\citenamefont {Langfeld}\ and\ \citenamefont
  {Pawlowski}(2013)}]{langfeld2013two}%
  \BibitemOpen
  \bibfield  {author} {\bibinfo {author} {\bibfnamefont {K.}~\bibnamefont
  {Langfeld}}\ and\ \bibinfo {author} {\bibfnamefont {J.~M.}\ \bibnamefont
  {Pawlowski}},\ }\href {https://doi.org/10.1103/PhysRevD.88.071502} {\bibfield
   {journal} {\bibinfo  {journal} {Phys. Rev. D}\ }\textbf {\bibinfo {volume}
  {88}},\ \bibinfo {pages} {071502} (\bibinfo {year} {2013})},\ \Eprint
  {https://arxiv.org/abs/1307.0455} {arXiv:1307.0455 [hep-lat]} \BibitemShut
  {NoStop}%
\bibitem [{\citenamefont {Langfeld}\ and\ \citenamefont
  {Lucini}(2014)}]{langfeld2014density}%
  \BibitemOpen
  \bibfield  {author} {\bibinfo {author} {\bibfnamefont {K.}~\bibnamefont
  {Langfeld}}\ and\ \bibinfo {author} {\bibfnamefont {B.}~\bibnamefont
  {Lucini}},\ }\href {https://doi.org/10.1103/PhysRevD.90.094502} {\bibfield
  {journal} {\bibinfo  {journal} {Phys. Rev. D}\ }\textbf {\bibinfo {volume}
  {90}},\ \bibinfo {pages} {094502} (\bibinfo {year} {2014})},\ \Eprint
  {https://arxiv.org/abs/1404.7187} {arXiv:1404.7187 [hep-lat]} \BibitemShut
  {NoStop}%
\bibitem [{\citenamefont {Gattringer}\ \emph {et~al.}(2019)\citenamefont
  {Gattringer}, \citenamefont {Mandl},\ and\ \citenamefont
  {T\"orek}}]{gattringer2019}%
  \BibitemOpen
  \bibfield  {author} {\bibinfo {author} {\bibfnamefont {C.}~\bibnamefont
  {Gattringer}}, \bibinfo {author} {\bibfnamefont {M.}~\bibnamefont {Mandl}},\
  and\ \bibinfo {author} {\bibfnamefont {P.}~\bibnamefont {T\"orek}},\ }\href
  {https://doi.org/10.1103/PhysRevD.100.114517} {\bibfield  {journal} {\bibinfo
   {journal} {Phys. Rev. D}\ }\textbf {\bibinfo {volume} {100}},\ \bibinfo
  {pages} {114517} (\bibinfo {year} {2019})}\BibitemShut {NoStop}%
\bibitem [{\citenamefont {Gattringer}\ \emph {et~al.}(2020)\citenamefont
  {Gattringer}, \citenamefont {Mandl},\ and\ \citenamefont
  {T\"orek}}]{Gattringer:2019egx}%
  \BibitemOpen
  \bibfield  {author} {\bibinfo {author} {\bibfnamefont {C.}~\bibnamefont
  {Gattringer}}, \bibinfo {author} {\bibfnamefont {M.}~\bibnamefont {Mandl}},\
  and\ \bibinfo {author} {\bibfnamefont {P.}~\bibnamefont {T\"orek}},\ }\href
  {https://doi.org/10.3390/particles3010008} {\bibfield  {journal} {\bibinfo
  {journal} {Particles}\ }\textbf {\bibinfo {volume} {3}},\ \bibinfo {pages}
  {87} (\bibinfo {year} {2020})},\ \Eprint {https://arxiv.org/abs/1912.05040}
  {arXiv:1912.05040 [hep-lat]} \BibitemShut {NoStop}%
\bibitem [{\citenamefont {Martinez}\ \emph {et~al.}(2016)\citenamefont
  {Martinez} \emph {et~al.}}]{martinez2016real}%
  \BibitemOpen
  \bibfield  {author} {\bibinfo {author} {\bibfnamefont {E.~A.}\ \bibnamefont
  {Martinez}} \emph {et~al.},\ }\href {https://doi.org/10.1038/nature18318}
  {\bibfield  {journal} {\bibinfo  {journal} {Nature}\ }\textbf {\bibinfo
  {volume} {534}},\ \bibinfo {pages} {516} (\bibinfo {year} {2016})},\ \Eprint
  {https://arxiv.org/abs/1605.04570} {arXiv:1605.04570 [quant-ph]} \BibitemShut
  {NoStop}%
\bibitem [{\citenamefont {Klco}\ \emph {et~al.}(2018)\citenamefont {Klco},
  \citenamefont {Dumitrescu}, \citenamefont {McCaskey}, \citenamefont {Morris},
  \citenamefont {Pooser}, \citenamefont {Sanz}, \citenamefont {Solano},
  \citenamefont {Lougovski},\ and\ \citenamefont
  {Savage}}]{PhysRevA.98.032331}%
  \BibitemOpen
  \bibfield  {author} {\bibinfo {author} {\bibfnamefont {N.}~\bibnamefont
  {Klco}}, \bibinfo {author} {\bibfnamefont {E.~F.}\ \bibnamefont
  {Dumitrescu}}, \bibinfo {author} {\bibfnamefont {A.~J.}\ \bibnamefont
  {McCaskey}}, \bibinfo {author} {\bibfnamefont {T.~D.}\ \bibnamefont
  {Morris}}, \bibinfo {author} {\bibfnamefont {R.~C.}\ \bibnamefont {Pooser}},
  \bibinfo {author} {\bibfnamefont {M.}~\bibnamefont {Sanz}}, \bibinfo {author}
  {\bibfnamefont {E.}~\bibnamefont {Solano}}, \bibinfo {author} {\bibfnamefont
  {P.}~\bibnamefont {Lougovski}},\ and\ \bibinfo {author} {\bibfnamefont
  {M.~J.}\ \bibnamefont {Savage}},\ }\href
  {https://doi.org/10.1103/PhysRevA.98.032331} {\bibfield  {journal} {\bibinfo
  {journal} {Phys. Rev. A}\ }\textbf {\bibinfo {volume} {98}},\ \bibinfo
  {pages} {032331} (\bibinfo {year} {2018})}\BibitemShut {NoStop}%
\bibitem [{\citenamefont {Yang}\ \emph
  {et~al.}(2020{\natexlab{b}})\citenamefont {Yang}, \citenamefont {Sun},
  \citenamefont {Ott}, \citenamefont {Wang}, \citenamefont {Zache},
  \citenamefont {Halimeh}, \citenamefont {Yuan}, \citenamefont {Hauke},\ and\
  \citenamefont {Pan}}]{Yang:2020yer}%
  \BibitemOpen
  \bibfield  {author} {\bibinfo {author} {\bibfnamefont {B.}~\bibnamefont
  {Yang}}, \bibinfo {author} {\bibfnamefont {H.}~\bibnamefont {Sun}}, \bibinfo
  {author} {\bibfnamefont {R.}~\bibnamefont {Ott}}, \bibinfo {author}
  {\bibfnamefont {H.-Y.}\ \bibnamefont {Wang}}, \bibinfo {author}
  {\bibfnamefont {T.~V.}\ \bibnamefont {Zache}}, \bibinfo {author}
  {\bibfnamefont {J.~C.}\ \bibnamefont {Halimeh}}, \bibinfo {author}
  {\bibfnamefont {Z.-S.}\ \bibnamefont {Yuan}}, \bibinfo {author}
  {\bibfnamefont {P.}~\bibnamefont {Hauke}},\ and\ \bibinfo {author}
  {\bibfnamefont {J.-W.}\ \bibnamefont {Pan}},\ }\href
  {https://doi.org/10.1038/s41586-020-2910-8} {\bibfield  {journal} {\bibinfo
  {journal} {Nature}\ }\textbf {\bibinfo {volume} {587}},\ \bibinfo {pages}
  {392} (\bibinfo {year} {2020}{\natexlab{b}})},\ \Eprint
  {https://arxiv.org/abs/2003.08945} {arXiv:2003.08945 [cond-mat.quant-gas]}
  \BibitemShut {NoStop}%
\bibitem [{\citenamefont {Ba\~nuls}\ and\ \citenamefont
  {Cichy}(2020)}]{banuls2020review}%
  \BibitemOpen
  \bibfield  {author} {\bibinfo {author} {\bibfnamefont {M.~C.}\ \bibnamefont
  {Ba\~nuls}}\ and\ \bibinfo {author} {\bibfnamefont {K.}~\bibnamefont
  {Cichy}},\ }\href {https://doi.org/10.1088/1361-6633/ab6311} {\bibfield
  {journal} {\bibinfo  {journal} {Rept. Prog. Phys.}\ }\textbf {\bibinfo
  {volume} {83}},\ \bibinfo {pages} {024401} (\bibinfo {year} {2020})},\
  \Eprint {https://arxiv.org/abs/1910.00257} {arXiv:1910.00257 [hep-lat]}
  \BibitemShut {NoStop}%
\bibitem [{\citenamefont {{Ba\~nuls, Mari Carmen}}\ \emph
  {et~al.}(2020)\citenamefont {{Ba\~nuls, Mari Carmen}}, \citenamefont {{Blatt,
  Rainer}}, \citenamefont {{Catani, Jacopo}}, \citenamefont {{Celi, Alessio}},
  \citenamefont {{Cirac, Juan Ignacio}}, \citenamefont {{Dalmonte, Marcello}},
  \citenamefont {{Fallani, Leonardo}}, \citenamefont {{Jansen, Karl}},
  \citenamefont {{Lewenstein, Maciej}}, \citenamefont {{Montangero, Simone}},
  \citenamefont {{Muschik, Christine A.}}, \citenamefont {{Reznik, Benni}},
  \citenamefont {{Rico, Enrique}}, \citenamefont {{Tagliacozzo, Luca}},
  \citenamefont {{Van Acoleyen, Karel}}, \citenamefont {{Verstraete, Frank}},
  \citenamefont {{Wiese, Uwe-Jens}}, \citenamefont {{Wingate, Matthew}},
  \citenamefont {{Zakrzewski, Jakub}},\ and\ \citenamefont {{Zoller,
  Peter}}}]{qtflag2020}%
  \BibitemOpen
  \bibfield  {author} {\bibinfo {author} {\bibnamefont {{Ba\~nuls, Mari
  Carmen}}}, \bibinfo {author} {\bibnamefont {{Blatt, Rainer}}}, \bibinfo
  {author} {\bibnamefont {{Catani, Jacopo}}}, \bibinfo {author} {\bibnamefont
  {{Celi, Alessio}}}, \bibinfo {author} {\bibnamefont {{Cirac, Juan Ignacio}}},
  \bibinfo {author} {\bibnamefont {{Dalmonte, Marcello}}}, \bibinfo {author}
  {\bibnamefont {{Fallani, Leonardo}}}, \bibinfo {author} {\bibnamefont
  {{Jansen, Karl}}}, \bibinfo {author} {\bibnamefont {{Lewenstein, Maciej}}},
  \bibinfo {author} {\bibnamefont {{Montangero, Simone}}}, \bibinfo {author}
  {\bibnamefont {{Muschik, Christine A.}}}, \bibinfo {author} {\bibnamefont
  {{Reznik, Benni}}}, \bibinfo {author} {\bibnamefont {{Rico, Enrique}}},
  \bibinfo {author} {\bibnamefont {{Tagliacozzo, Luca}}}, \bibinfo {author}
  {\bibnamefont {{Van Acoleyen, Karel}}}, \bibinfo {author} {\bibnamefont
  {{Verstraete, Frank}}}, \bibinfo {author} {\bibnamefont {{Wiese, Uwe-Jens}}},
  \bibinfo {author} {\bibnamefont {{Wingate, Matthew}}}, \bibinfo {author}
  {\bibnamefont {{Zakrzewski, Jakub}}},\ and\ \bibinfo {author} {\bibnamefont
  {{Zoller, Peter}}},\ }\href {https://doi.org/10.1140/epjd/e2020-100571-8}
  {\bibfield  {journal} {\bibinfo  {journal} {Eur. Phys. J. D}\ }\textbf
  {\bibinfo {volume} {74}},\ \bibinfo {pages} {165} (\bibinfo {year}
  {2020})}\BibitemShut {NoStop}%
\bibitem [{\citenamefont {Elser}\ and\ \citenamefont
  {Kalloniatis}(1996)}]{elser1996qed}%
  \BibitemOpen
  \bibfield  {author} {\bibinfo {author} {\bibfnamefont {S.}~\bibnamefont
  {Elser}}\ and\ \bibinfo {author} {\bibfnamefont {A.~C.}\ \bibnamefont
  {Kalloniatis}},\ }\href {https://doi.org/10.1016/0370-2693(96)00201-8}
  {\bibfield  {journal} {\bibinfo  {journal} {Phys. Lett. B}\ }\textbf
  {\bibinfo {volume} {375}},\ \bibinfo {pages} {285} (\bibinfo {year}
  {1996})},\ \Eprint {https://arxiv.org/abs/hep-th/9601045}
  {arXiv:hep-th/9601045} \BibitemShut {NoStop}%
\bibitem [{\citenamefont {Schwinger}(1962)}]{schwinger1962gauge}%
  \BibitemOpen
  \bibfield  {author} {\bibinfo {author} {\bibfnamefont {J.}~\bibnamefont
  {Schwinger}},\ }\href {https://doi.org/10.1103/PhysRev.128.2425} {\bibfield
  {journal} {\bibinfo  {journal} {Phys. Rev.}\ }\textbf {\bibinfo {volume}
  {128}},\ \bibinfo {pages} {2425} (\bibinfo {year} {1962})}\BibitemShut
  {NoStop}%
\bibitem [{\citenamefont {Coleman}(1976)}]{coleman1976more}%
  \BibitemOpen
  \bibfield  {author} {\bibinfo {author} {\bibfnamefont {S.~R.}\ \bibnamefont
  {Coleman}},\ }\href {https://doi.org/10.1016/0003-4916(76)90280-3} {\bibfield
   {journal} {\bibinfo  {journal} {Annals Phys.}\ }\textbf {\bibinfo {volume}
  {101}},\ \bibinfo {pages} {239} (\bibinfo {year} {1976})}\BibitemShut
  {NoStop}%
\bibitem [{\citenamefont {Byrnes}\ \emph {et~al.}(2002)\citenamefont {Byrnes},
  \citenamefont {Sriganesh}, \citenamefont {Bursill},\ and\ \citenamefont
  {Hamer}}]{Byrnes:2002nv}%
  \BibitemOpen
  \bibfield  {author} {\bibinfo {author} {\bibfnamefont {T.}~\bibnamefont
  {Byrnes}}, \bibinfo {author} {\bibfnamefont {P.}~\bibnamefont {Sriganesh}},
  \bibinfo {author} {\bibfnamefont {R.~J.}\ \bibnamefont {Bursill}},\ and\
  \bibinfo {author} {\bibfnamefont {C.~J.}\ \bibnamefont {Hamer}},\ }\href
  {https://doi.org/10.1103/PhysRevD.66.013002} {\bibfield  {journal} {\bibinfo
  {journal} {Phys. Rev. D}\ }\textbf {\bibinfo {volume} {66}},\ \bibinfo
  {pages} {013002} (\bibinfo {year} {2002})},\ \Eprint
  {https://arxiv.org/abs/hep-lat/0202014} {arXiv:hep-lat/0202014} \BibitemShut
  {NoStop}%
\bibitem [{\citenamefont {Kogut}\ and\ \citenamefont
  {Susskind}(1975)}]{kogut1975j}%
  \BibitemOpen
  \bibfield  {author} {\bibinfo {author} {\bibfnamefont {J.}~\bibnamefont
  {Kogut}}\ and\ \bibinfo {author} {\bibfnamefont {L.}~\bibnamefont
  {Susskind}},\ }\href {https://doi.org/10.1103/PhysRevD.11.395} {\bibfield
  {journal} {\bibinfo  {journal} {Phys. Rev. D}\ }\textbf {\bibinfo {volume}
  {11}},\ \bibinfo {pages} {395} (\bibinfo {year} {1975})}\BibitemShut
  {NoStop}%
\bibitem [{\citenamefont {Hamer}\ \emph {et~al.}(1997)\citenamefont {Hamer},
  \citenamefont {Zheng},\ and\ \citenamefont {Oitmaa}}]{hamer1997series}%
  \BibitemOpen
  \bibfield  {author} {\bibinfo {author} {\bibfnamefont {C.~J.}\ \bibnamefont
  {Hamer}}, \bibinfo {author} {\bibfnamefont {W.-h.}\ \bibnamefont {Zheng}},\
  and\ \bibinfo {author} {\bibfnamefont {J.}~\bibnamefont {Oitmaa}},\ }\href
  {https://doi.org/10.1103/PhysRevD.56.55} {\bibfield  {journal} {\bibinfo
  {journal} {Phys. Rev. D}\ }\textbf {\bibinfo {volume} {56}},\ \bibinfo
  {pages} {55} (\bibinfo {year} {1997})},\ \Eprint
  {https://arxiv.org/abs/hep-lat/9701015} {arXiv:hep-lat/9701015} \BibitemShut
  {NoStop}%
\bibitem [{\citenamefont {Banks}\ \emph {et~al.}(1976)\citenamefont {Banks},
  \citenamefont {Susskind},\ and\ \citenamefont {Kogut}}]{banks1976strong}%
  \BibitemOpen
  \bibfield  {author} {\bibinfo {author} {\bibfnamefont {T.}~\bibnamefont
  {Banks}}, \bibinfo {author} {\bibfnamefont {L.}~\bibnamefont {Susskind}},\
  and\ \bibinfo {author} {\bibfnamefont {J.}~\bibnamefont {Kogut}},\ }\href
  {https://doi.org/10.1103/PhysRevD.13.1043} {\bibfield  {journal} {\bibinfo
  {journal} {Phys. Rev. D}\ }\textbf {\bibinfo {volume} {13}},\ \bibinfo
  {pages} {1043} (\bibinfo {year} {1976})}\BibitemShut {NoStop}%
\bibitem [{\citenamefont {Holzner}\ \emph {et~al.}(2011)\citenamefont
  {Holzner}, \citenamefont {Weichselbaum}, \citenamefont {McCulloch},
  \citenamefont {Schollw\"ock},\ and\ \citenamefont {von
  Delft}}]{PhysRevB.83.195115}%
  \BibitemOpen
  \bibfield  {author} {\bibinfo {author} {\bibfnamefont {A.}~\bibnamefont
  {Holzner}}, \bibinfo {author} {\bibfnamefont {A.}~\bibnamefont
  {Weichselbaum}}, \bibinfo {author} {\bibfnamefont {I.~P.}\ \bibnamefont
  {McCulloch}}, \bibinfo {author} {\bibfnamefont {U.}~\bibnamefont
  {Schollw\"ock}},\ and\ \bibinfo {author} {\bibfnamefont {J.}~\bibnamefont
  {von Delft}},\ }\href {https://doi.org/10.1103/PhysRevB.83.195115} {\bibfield
   {journal} {\bibinfo  {journal} {Phys. Rev. B}\ }\textbf {\bibinfo {volume}
  {83}},\ \bibinfo {pages} {195115} (\bibinfo {year} {2011})}\BibitemShut
  {NoStop}%
\bibitem [{\citenamefont {Wolf}\ \emph {et~al.}(2015)\citenamefont {Wolf},
  \citenamefont {Justiniano}, \citenamefont {McCulloch},\ and\ \citenamefont
  {Schollw\"ock}}]{Wolf2015}%
  \BibitemOpen
  \bibfield  {author} {\bibinfo {author} {\bibfnamefont {F.~A.}\ \bibnamefont
  {Wolf}}, \bibinfo {author} {\bibfnamefont {J.~A.}\ \bibnamefont
  {Justiniano}}, \bibinfo {author} {\bibfnamefont {I.~P.}\ \bibnamefont
  {McCulloch}},\ and\ \bibinfo {author} {\bibfnamefont {U.}~\bibnamefont
  {Schollw\"ock}},\ }\href {https://doi.org/10.1103/PhysRevB.91.115144}
  {\bibfield  {journal} {\bibinfo  {journal} {Phys. Rev. B}\ }\textbf {\bibinfo
  {volume} {91}},\ \bibinfo {pages} {115144} (\bibinfo {year}
  {2015})}\BibitemShut {NoStop}%
\bibitem [{\citenamefont {Halimeh}\ \emph {et~al.}(2015)\citenamefont
  {Halimeh}, \citenamefont {Kolley},\ and\ \citenamefont
  {McCulloch}}]{Halimeh2015}%
  \BibitemOpen
  \bibfield  {author} {\bibinfo {author} {\bibfnamefont {J.~C.}\ \bibnamefont
  {Halimeh}}, \bibinfo {author} {\bibfnamefont {F.}~\bibnamefont {Kolley}},\
  and\ \bibinfo {author} {\bibfnamefont {I.~P.}\ \bibnamefont {McCulloch}},\
  }\href {https://doi.org/10.1103/PhysRevB.92.115130} {\bibfield  {journal}
  {\bibinfo  {journal} {Phys. Rev. B}\ }\textbf {\bibinfo {volume} {92}},\
  \bibinfo {pages} {115130} (\bibinfo {year} {2015})}\BibitemShut {NoStop}%
\bibitem [{\citenamefont {Verstraete}\ \emph
  {et~al.}(2004{\natexlab{b}})\citenamefont {Verstraete}, \citenamefont
  {Garc\'{\i}a-Ripoll},\ and\ \citenamefont {Cirac}}]{PhysRevLett.93.207204}%
  \BibitemOpen
  \bibfield  {author} {\bibinfo {author} {\bibfnamefont {F.}~\bibnamefont
  {Verstraete}}, \bibinfo {author} {\bibfnamefont {J.~J.}\ \bibnamefont
  {Garc\'{\i}a-Ripoll}},\ and\ \bibinfo {author} {\bibfnamefont {J.~I.}\
  \bibnamefont {Cirac}},\ }\href
  {https://doi.org/10.1103/PhysRevLett.93.207204} {\bibfield  {journal}
  {\bibinfo  {journal} {Phys. Rev. Lett.}\ }\textbf {\bibinfo {volume} {93}},\
  \bibinfo {pages} {207204} (\bibinfo {year} {2004}{\natexlab{b}})}\BibitemShut
  {NoStop}%
\bibitem [{\citenamefont {Moore}(2011)}]{MOORE2011537}%
  \BibitemOpen
  \bibfield  {author} {\bibinfo {author} {\bibfnamefont {G.}~\bibnamefont
  {Moore}},\ }\href {https://doi.org/https://doi.org/10.1016/j.laa.2010.09.021}
  {\bibfield  {journal} {\bibinfo  {journal} {Linear Algebra and its
  Applications}\ }\textbf {\bibinfo {volume} {435}},\ \bibinfo {pages} {537 }
  (\bibinfo {year} {2011})},\ \bibinfo {note} {special Issue: Dedication to
  Pete Stewart on the occasion of his 70th birthday}\BibitemShut {NoStop}%
\bibitem [{\citenamefont {Ba\~nuls}\ \emph {et~al.}(2013)\citenamefont
  {Ba\~nuls}, \citenamefont {Cichy}, \citenamefont {Jansen},\ and\
  \citenamefont {Cirac}}]{Banuls:2013ja}%
  \BibitemOpen
  \bibfield  {author} {\bibinfo {author} {\bibfnamefont {M.~C.}\ \bibnamefont
  {Ba\~nuls}}, \bibinfo {author} {\bibfnamefont {K.}~\bibnamefont {Cichy}},
  \bibinfo {author} {\bibfnamefont {K.}~\bibnamefont {Jansen}},\ and\ \bibinfo
  {author} {\bibfnamefont {J.~I.}\ \bibnamefont {Cirac}},\ }\href
  {https://doi.org/10.1007/JHEP11(2013)158} {\bibfield  {journal} {\bibinfo
  {journal} {JHEP}\ }\textbf {\bibinfo {volume} {11}},\ \bibinfo {pages}
  {158}},\ \Eprint {https://arxiv.org/abs/1305.3765} {arXiv:1305.3765
  [hep-lat]} \BibitemShut {NoStop}%
\bibitem [{\citenamefont {Pauli}\ and\ \citenamefont
  {Brodsky}(1985)}]{pauli1993phys}%
  \BibitemOpen
  \bibfield  {author} {\bibinfo {author} {\bibfnamefont {H.-C.}\ \bibnamefont
  {Pauli}}\ and\ \bibinfo {author} {\bibfnamefont {S.~J.}\ \bibnamefont
  {Brodsky}},\ }\href {https://doi.org/10.1103/PhysRevD.32.1993} {\bibfield
  {journal} {\bibinfo  {journal} {Phys. Rev. D}\ }\textbf {\bibinfo {volume}
  {32}},\ \bibinfo {pages} {1993} (\bibinfo {year} {1985})}\BibitemShut
  {NoStop}%
\bibitem [{\citenamefont {Brodsky}\ \emph {et~al.}(1993)\citenamefont
  {Brodsky}, \citenamefont {McCartor}, \citenamefont {Pauli},\ and\
  \citenamefont {Pinsky}}]{brodsky1993slac}%
  \BibitemOpen
  \bibfield  {author} {\bibinfo {author} {\bibfnamefont {S.}~\bibnamefont
  {Brodsky}}, \bibinfo {author} {\bibfnamefont {G.}~\bibnamefont {McCartor}},
  \bibinfo {author} {\bibfnamefont {H.}~\bibnamefont {Pauli}},\ and\ \bibinfo
  {author} {\bibfnamefont {S.}~\bibnamefont {Pinsky}},\ }\href@noop {}
  {\bibfield  {journal} {\bibinfo  {journal} {Particle World}\ }\textbf
  {\bibinfo {volume} {3}},\ \bibinfo {pages} {109} (\bibinfo {year}
  {1993})}\BibitemShut {NoStop}%
\bibitem [{\citenamefont {Brodsky}\ \emph {et~al.}(1998)\citenamefont
  {Brodsky}, \citenamefont {Pauli},\ and\ \citenamefont
  {Pinsky}}]{brodsky1998quantum}%
  \BibitemOpen
  \bibfield  {author} {\bibinfo {author} {\bibfnamefont {S.~J.}\ \bibnamefont
  {Brodsky}}, \bibinfo {author} {\bibfnamefont {H.-C.}\ \bibnamefont {Pauli}},\
  and\ \bibinfo {author} {\bibfnamefont {S.~S.}\ \bibnamefont {Pinsky}},\
  }\href {https://doi.org/10.1016/S0370-1573(97)00089-6} {\bibfield  {journal}
  {\bibinfo  {journal} {Phys. Rept.}\ }\textbf {\bibinfo {volume} {301}},\
  \bibinfo {pages} {299} (\bibinfo {year} {1998})},\ \Eprint
  {https://arxiv.org/abs/hep-ph/9705477} {arXiv:hep-ph/9705477} \BibitemShut
  {NoStop}%
\bibitem [{\citenamefont {{Hartmann}}\ \emph {et~al.}(2005)\citenamefont
  {{Hartmann}}, \citenamefont {{Mahler}},\ and\ \citenamefont
  {{Hess}}}]{hartmann2005}%
  \BibitemOpen
  \bibfield  {author} {\bibinfo {author} {\bibfnamefont {M.}~\bibnamefont
  {{Hartmann}}}, \bibinfo {author} {\bibfnamefont {G.}~\bibnamefont
  {{Mahler}}},\ and\ \bibinfo {author} {\bibfnamefont {O.}~\bibnamefont
  {{Hess}}},\ }\href {https://doi.org/10.1007/s10955-004-4298-5} {\bibfield
  {journal} {\bibinfo  {journal} {Journal of Statistical Physics}\ }\textbf
  {\bibinfo {volume} {119}},\ \bibinfo {pages} {1139} (\bibinfo {year}
  {2005})},\ \Eprint {https://arxiv.org/abs/cond-mat/0406100}
  {arXiv:cond-mat/0406100 [cond-mat.stat-mech]} \BibitemShut {NoStop}%
\bibitem [{\citenamefont {{Keating}}\ \emph {et~al.}(2015)\citenamefont
  {{Keating}}, \citenamefont {{Linden}},\ and\ \citenamefont
  {{Wells}}}]{keating2015}%
  \BibitemOpen
  \bibfield  {author} {\bibinfo {author} {\bibfnamefont {J.~P.}\ \bibnamefont
  {{Keating}}}, \bibinfo {author} {\bibfnamefont {N.}~\bibnamefont
  {{Linden}}},\ and\ \bibinfo {author} {\bibfnamefont {H.~J.}\ \bibnamefont
  {{Wells}}},\ }\href {https://doi.org/10.1007/s00220-015-2366-0} {\bibfield
  {journal} {\bibinfo  {journal} {Communications in Mathematical Physics}\
  }\textbf {\bibinfo {volume} {338}},\ \bibinfo {pages} {81} (\bibinfo {year}
  {2015})},\ \Eprint {https://arxiv.org/abs/1403.1121} {arXiv:1403.1121
  [math-ph]} \BibitemShut {NoStop}%
\bibitem [{Note1()}]{Note1}%
  \BibitemOpen
  \bibinfo {note} {Notice that in~\cite {elser1996qed} the Gaussian behaviour
  of the DOS was only a qualitative observation, since the parameters of the
  numerical study did not allow approaching the continuum.}\BibitemShut {Stop}%
\bibitem [{\citenamefont {Ba\~nuls}\ \emph {et~al.}(2015)\citenamefont
  {Ba\~nuls}, \citenamefont {Cichy}, \citenamefont {Cirac}, \citenamefont
  {Jansen},\ and\ \citenamefont {Saito}}]{banuls2015thermal}%
  \BibitemOpen
  \bibfield  {author} {\bibinfo {author} {\bibfnamefont {M.~C.}\ \bibnamefont
  {Ba\~nuls}}, \bibinfo {author} {\bibfnamefont {K.}~\bibnamefont {Cichy}},
  \bibinfo {author} {\bibfnamefont {J.~I.}\ \bibnamefont {Cirac}}, \bibinfo
  {author} {\bibfnamefont {K.}~\bibnamefont {Jansen}},\ and\ \bibinfo {author}
  {\bibfnamefont {H.}~\bibnamefont {Saito}},\ }\href
  {https://doi.org/10.1103/PhysRevD.92.034519} {\bibfield  {journal} {\bibinfo
  {journal} {Phys. Rev. D}\ }\textbf {\bibinfo {volume} {92}},\ \bibinfo
  {pages} {034519} (\bibinfo {year} {2015})},\ \Eprint
  {https://arxiv.org/abs/1505.00279} {arXiv:1505.00279 [hep-lat]} \BibitemShut
  {NoStop}%
\bibitem [{\citenamefont {Sachs}\ and\ \citenamefont
  {Wipf}(1992)}]{Sachs:1991en}%
  \BibitemOpen
  \bibfield  {author} {\bibinfo {author} {\bibfnamefont {I.}~\bibnamefont
  {Sachs}}\ and\ \bibinfo {author} {\bibfnamefont {A.}~\bibnamefont {Wipf}},\
  }\href@noop {} {\bibfield  {journal} {\bibinfo  {journal} {Helv. Phys. Acta}\
  }\textbf {\bibinfo {volume} {65}},\ \bibinfo {pages} {652} (\bibinfo {year}
  {1992})},\ \Eprint {https://arxiv.org/abs/1005.1822} {arXiv:1005.1822
  [hep-th]} \BibitemShut {NoStop}%
\bibitem [{\citenamefont {Ba\~nuls}\ \emph {et~al.}(2016)\citenamefont
  {Ba\~nuls}, \citenamefont {Cichy}, \citenamefont {Jansen},\ and\
  \citenamefont {Saito}}]{banuls2016thermal}%
  \BibitemOpen
  \bibfield  {author} {\bibinfo {author} {\bibfnamefont {M.~C.}\ \bibnamefont
  {Ba\~nuls}}, \bibinfo {author} {\bibfnamefont {K.}~\bibnamefont {Cichy}},
  \bibinfo {author} {\bibfnamefont {K.}~\bibnamefont {Jansen}},\ and\ \bibinfo
  {author} {\bibfnamefont {H.}~\bibnamefont {Saito}},\ }\href
  {https://doi.org/10.1103/PhysRevD.93.094512} {\bibfield  {journal} {\bibinfo
  {journal} {Phys. Rev. D}\ }\textbf {\bibinfo {volume} {93}},\ \bibinfo
  {pages} {094512} (\bibinfo {year} {2016})},\ \Eprint
  {https://arxiv.org/abs/1603.05002} {arXiv:1603.05002 [hep-lat]} \BibitemShut
  {NoStop}%
\bibitem [{\citenamefont {Buyens}\ \emph {et~al.}(2016)\citenamefont {Buyens},
  \citenamefont {Verstraete},\ and\ \citenamefont
  {Van~Acoleyen}}]{Buyens:2016ecr}%
  \BibitemOpen
  \bibfield  {author} {\bibinfo {author} {\bibfnamefont {B.}~\bibnamefont
  {Buyens}}, \bibinfo {author} {\bibfnamefont {F.}~\bibnamefont {Verstraete}},\
  and\ \bibinfo {author} {\bibfnamefont {K.}~\bibnamefont {Van~Acoleyen}},\
  }\href {https://doi.org/10.1103/PhysRevD.94.085018} {\bibfield  {journal}
  {\bibinfo  {journal} {Phys. Rev. D}\ }\textbf {\bibinfo {volume} {94}},\
  \bibinfo {pages} {085018} (\bibinfo {year} {2016})},\ \Eprint
  {https://arxiv.org/abs/1606.03385} {arXiv:1606.03385 [hep-lat]} \BibitemShut
  {NoStop}%
\bibitem [{Note2()}]{Note2}%
  \BibitemOpen
  \bibinfo {note} {Specifically, for $\beta _{0}=0.1$ and the system with
  $N=20$ and $x=5$, the Bessel function of order 10 has magnitude
  $|I_{10}(-\beta _{0}/\alpha )|\sim 7\times 10^{15}$, whereas for the system
  $N=40$ and $x=5$, $|I_{10}(-\beta _{0}/\alpha )|\sim 8\times
  10^{129}$.}\BibitemShut {Stop}%
\bibitem [{Note3()}]{Note3}%
  \BibitemOpen
  \bibinfo {note} {Even though the standard MPO method, which proceeds via
  imaginary time evolution, becomes less accurate as $\beta $ increases due to
  accumulation of truncation, it is possible to estimate this error
  systematically (see~\cite {banuls2015thermal}) and use sufficiently converged
  values as reference, as is done in Fig~\ref {fig:
  condensateN20N30}.}\BibitemShut {Stop}%
\bibitem [{\citenamefont {Strauss}\ and\ \citenamefont
  {Beyer}(2008)}]{Strauss2008prl}%
  \BibitemOpen
  \bibfield  {author} {\bibinfo {author} {\bibfnamefont {S.}~\bibnamefont
  {Strauss}}\ and\ \bibinfo {author} {\bibfnamefont {M.}~\bibnamefont
  {Beyer}},\ }\href {https://doi.org/10.1103/PhysRevLett.101.100402} {\bibfield
   {journal} {\bibinfo  {journal} {Phys. Rev. Lett.}\ }\textbf {\bibinfo
  {volume} {101}},\ \bibinfo {pages} {100402} (\bibinfo {year}
  {2008})}\BibitemShut {NoStop}%
\bibitem [{\citenamefont {Strauss}\ and\ \citenamefont
  {Beyer}(2009)}]{Strauss2009pos}%
  \BibitemOpen
  \bibfield  {author} {\bibinfo {author} {\bibfnamefont {S.}~\bibnamefont
  {Strauss}}\ and\ \bibinfo {author} {\bibfnamefont {M.}~\bibnamefont
  {Beyer}},\ }\href {https://doi.org/10.22323/1.061.0010} {\bibfield  {journal}
  {\bibinfo  {journal} {PoS}\ }\textbf {\bibinfo {volume} {LC2008}},\ \bibinfo
  {pages} {010} (\bibinfo {year} {2009})}\BibitemShut {NoStop}%
\end{thebibliography}%


%
\end{document}